\documentclass[12pt]{iopart}
\pdfoutput=1
\usepackage{graphicx,sidecap}
\usepackage{bm}
\usepackage{braket}
\usepackage{amssymb}
\usepackage{mathrsfs}
\usepackage{amsmath}
\usepackage{xcolor}
\usepackage{hyperref}
\usepackage{enumerate}
\usepackage[toc,title]{appendix}
\usepackage{verbatim}
\usepackage[margin=30pt, bf, font=footnotesize, center, justification=justified]{caption}[2004/07/16]
\hypersetup{colorlinks,bookmarksopen,bookmarksnumbered,
 citecolor=red,
 linkcolor=blue,
 pdfstartview=green,
 urlcolor=purple}
 
\catcode`@=11 
 \renewcommand\tableofcontents{%
   \section*{\contentsname}%
   \@starttoc{toc}%
}
\catcode`@=12

\def\be{\begin{equation}}
\def\ee{\end{equation}}
\def\bea{\begin{eqnarray}}
\def\eea{\end{eqnarray}}
\def\Tr{{\rm Tr}}
\def\iu{\textrm{i}}
\def\ket|#1>{| #1 \rangle}
\def\bra<#1|{\langle #1 |}
\def\<{\langle}
\def\>{\rangle}
\def\{{\lbrace}
\def\}{\rbrace}
\def\({\left(}
\def\){\right)}
\def\bbeta{\boldsymbol{\beta}}

\def\barray{\begin{eqnarray}}
\def\earray{\end{eqnarray}}
\def\beq{\begin{equation}}
\def\eeq{\end{equation}}

\def\Rmath{\mathbb{R}}

\def\INT{\bm{\mathfrak{S}}}
\def\INT{\mathbb{S}}
\def\D{\mathbb{D}}
\def\AN{\mathbb{A}} 

\begin{document}

\title[Entanglement hamiltonian and contour in inhomogeneous 1D critical  systems]{Entanglement hamiltonian and entanglement contour in inhomogeneous 1D critical systems}

\author{Erik Tonni$^1$, Javier Rodr\'{\i}guez-Laguna$^2$ and Germ\'an Sierra$^3$}

\address{$^1$\, SISSA and INFN, via Bonomea 265, 34136 Trieste, Italy.} 
\address{$^2$\, Departamento de F\'{\i}sica Fundamental, UNED, Madrid, Spain.}
\address{$^3$\, Instituto de F\'{\i}sica Te\'orica, UAM/CSIC, Madrid, Spain.}

\begin{abstract}
Inhomogeneous quantum critical systems in one spatial dimension have been studied 
by using  conformal field theory  in  static curved backgrounds. 
Two interesting  examples are the free fermion gas in the harmonic trap and 
the inhomogeneous XX spin chain called rainbow chain.
For conformal field theories defined on static curved spacetimes
characterised by a metric which is Weyl equivalent to the flat metric,
with the Weyl factor depending only on the spatial coordinate,
we study the entanglement hamiltonian and the entanglement spectrum
of an interval adjacent to the boundary of a segment 
where the same boundary condition is imposed at the endpoints. 
A contour function for the entanglement entropies 
corresponding to this configuration is also considered,
being closely related  to the entanglement hamiltonian. 
The analytic expressions obtained by considering
the curved spacetime which characterises the rainbow model
have been checked against numerical data for the rainbow chain, finding an excellent agreement.

\end{abstract}

\maketitle
\tableofcontents


\section{Introduction}
\label{sec:intro}

Entanglement has become a central tool to study extended quantum systems in different areas of theoretical physics
such as condensed matter physics, quantum information theory, quantum optics and quantum gravity \cite{reviews}. 
Recent important advances have allowed to set up experiments which have detected characteristic features of entanglement \cite{experiment}.

Entanglement in spatially inhomogeneous systems has attracted a lot of attention. 
Indeed, being inhomogeneity ubiquitous in experimental settings, 
it is important to assess the validity of the predicted universal features of
entanglement in inhomogeneous systems. 
For instance, quenched disorder in the coupling constants of a critical system in one spatial dimension
can lead to a behavior which is remarkably similar to the conformal
case, with a logarithmic growth of the average block entropies
\cite{rm-04, disorder chain}. 
Other interesting inhomogeneous systems in one spatial dimension 
are  fermionic systems in the presence of a trapping potential \cite{trap-pot, Dubail.17},
spin chains with gradients \cite{eisler-17}, 
spin chains with exponential growth of the couplings near the edges in order to reduce the boundary effects \cite{Ueda} 
and the {\em rainbow chain}, 
where exponentially reduced couplings towards the endpoints of a segment 
lead to a volumetric growth of the entanglement entropy in the ground state
\cite{Vitagliano.10,Ramirez.14b,Ramirez.15,Laguna.16,Laguna.17}.

The rainbow chain can be considered as the vacuum of a quantum field
theory on a curved background with negative curvature \cite{Laguna.17}.
Interestingly, the ground state of the  rainbow chain in a particular limit 
resembles a {\em thermofield double} state, 
i.e. each half of the chain behaves like a homogeneous system at a
finite temperature  \cite{Ramirez.15}.

 Other interesting models have been considered in \cite{ryu-ludwig-16}.
The similarity between smoothly inhomogeneous couplings 
and the occurrence of a static curved background metric
 has been employed to design simulators for effects which are
 characteristic  of quantum field theory in curved spacetimes, 
 such as the Unruh effect \cite{Boada-Laguna}.

 The entanglement entropies, i.e. the entanglement entropy and the R\'enyi entropies,
 are important quantities to study in order to  quantify the bipartite entanglement \cite{CW94, H94, V03, CC04}.
 Other quantities can be introduced which are expected to provide more information about the entanglement of a bipartition.
 In particular, in this manuscript we consider the entanglement hamiltonian \cite{bw, ent-ham, ent-ham-latt, Cardy-Tonni16, peschel-eisler-17}, 
 an operator whose spectrum gives the entanglement entropy and the R\'enyi entropies, 
 and the contour for the entanglement entropies \cite{br-04, chen-vidal, frerot-roschilde, cdt-17}.
 Our goal is to study quantitatively these magnitudes for some inhomogeneous critical systems in 
 one spatial dimension. 
 The rainbow model is the main benchmark of our analysis.

Given a quantum system in its ground state $\ket|\Psi>$ and a spatial
bipartition of the system into two subsystems $A$ and $B$ such that $A\cup B$ is the
entire space, one can assume that, correspondingly, the Hilbert
space of the system can be factorised as $\mathcal{H} = \mathcal{H}_A \otimes \mathcal{H}_B$. 
The reduced density matrix $\rho_A \equiv \Tr_B \ket|\Psi>\bra<\Psi|$ associated to the subsystem $A$ is obtained by tracing over the
degrees of freedom of $B$ and it can always be written as
\beq
\rho_A = e^{-2\pi K_A} \,,
\label{1}
\eeq 
where the operator $K_A$ is the entanglement hamiltonian. 
After the seminal work of Bisognano and Wichmann
\cite{bw}, several interesting quantitative results have been obtained about the explicit form of $K_A$ 
in quantum field theories \cite{ent-ham, Cardy-Tonni16}.
The entanglement hamiltonian has been studied also in some lattice models \cite{ent-ham-latt, peschel-eisler-17}.
The entanglement spectrum is given by the eigenvalues of $\rho_A$ \cite{single-copy-ent, ent-spec}.

The R\'enyi entropies are important scalar quantities defined from the $n$-th power of the reduced density matrix as follows
\beq
S^{(n)}_A \,=\, \frac{1}{1-n} \,\log {\rm Tr} \rho^n_A\,,
\label{2}
\eeq 
where $n \geqslant 2$ is an integer parameter. 
The entanglement entropy can be found from the R\'enyi entropies \eqref{2}
through the replica limit $S_A= \lim_{n \rightarrow 1} S^{(n)}_A$,
which requires to perform an analytic continuation of the integer
parameter $n$. 
In the following we denote by $S^{(n)}_A$ with $n\geqslant 1$ the entanglement entropies, 
meaning that $S^{(1)}_A \equiv S_A$. 
An important property to remind is $S^{(n)}_A = S^{(n)}_B$ for any $n\geqslant 1$
when the entire system is in a pure state.
It is worth remarking that the entanglement entropies are scalar quantities which 
can be constructed from the entanglement spectrum.

Another interesting quantity to consider is the contour for the
entanglement entropies \cite{br-04, chen-vidal, frerot-roschilde, cdt-17}. 
In a lattice model where a spatial bipartition
has been introduced, the contour for the entanglement entropies is
given by a non negative function $s_A^{(n)}(i)$ (called contour
function in the following) which provides information about the
contribution of the $i$-th site in $A$ to the entanglement between $A$
and $B$. The minimal properties that the contour function 
 must satisfy are
\be
\label{contour def lattice}
S_A^{(n)} = \sum_{i \,\in\,A} s_A^{(n)}(i)\,,
\;\;\qquad\;\;
s_A^{(n)}(i) \geqslant 0\,.
\ee

It is straightforward to observe that these two conditions do not
define the contour function $s_A^{(n)}(i) $ in a unique way. The
simplest function fulfilling \eqref{contour def lattice} is the flat contour
$s_A^{(n)}(i) = S_A^{(n)}/|A|$, where $|A|$ is the total number of
sites in $A$.
Nonetheless, we do not expect that all the sites in $A$ equally
contribute to $S_A^{(n)}$. Indeed, the main contribution should come
from the sites near the hypersurface separating $A$ and $B$
(often called entangling hypersurface). In order to improve the definition
of the contour for the entanglement entropies, further requirements
have been introduced in \cite{chen-vidal}, where an explicit
construction of $s_A^{(n)}(i)$ for free fermions on the lattice has been
proposed. In harmonic lattices, a contour function $s_A^{(n)}(i)$
satisfying \eqref{contour def lattice} and other properties close to
the ones introduced in \cite{chen-vidal} has been constructed in
\cite{cdt-17} (see also \cite{br-04, frerot-roschilde} for previous
work). 
The CFT description of the contour for the entanglement entropies
has been discussed in \cite{cdt-17}, by employing the analysis of \cite{Cardy-Tonni16}.
We remark that, while the entanglement spectrum provides all the
entanglement, further information is required to construct the contour for the
entanglement entropies.

In this manuscript we consider quantum systems on a one dimensional (1D)
lattice whose continuum limit is described by conformal field theories
(CFTs). 
Thus, the complementary domains $A$ and $B$ are made by intervals 
and the entangling hypersurface by isolated points. 
In $1+1$ dimensional quantum field theories $\Tr \rho_A^n$ can
be computed through the replica construction by considering $n$ copies
of the underlying model and joining them cyclically in a proper way.
An ultraviolet (UV) cutoff $\epsilon$ is needed to regularize the
ultraviolet divergencies. 
Following \cite{H94}, in the  two dimensional Euclidean spacetime 
where the field theory is defined, the UV cutoff can be introduced by
removing infinitesimal disks of radius $\epsilon$ centered at the
entangling points separating $A$ and $B$. 
In \cite{Cardy-Tonni16} this regularization procedure has been adopted to study the entanglement
hamiltonian $K_A$ for certain configurations.
This approach is adopted also in our analysis. 

Alternatively, $\Tr \rho_A^n$ can be computed as correlation functions
of twist fields located at the entangling points \cite{CC04, cardy-doyon}. 
In this approach the UV cutoff is introduced through a
thin slit of width $\epsilon$ separating the upper and the lower edge
of $A$ in the direction of the euclidean time. 
This method has been employed also for the entanglement entropies
of subsystems made by disjoint intervals \cite{2int-refs}.

In this manuscript we study the entanglement hamiltonian and 
the contour function for the entanglement entropies 
corresponding to a particular configuration 
in 1D inhomogeneous critical systems.
In particular, the system is defined on a segment where the same boundary condition
is imposed at the endpoints and the subsystem $A$ is an interval adjacent to 
one of the boundaries.
Analytic results are obtained which are valid for a large class of 
inhomogeneous critical systems in 1D.
The benchmark for our analytic expressions is the rainbow model.
A numerical analysis is performed for this inhomogeneous critical chain 
by adapting the method of \cite{Fujita.17} for the entanglement hamiltonian
and the method of \cite{chen-vidal} for the contour function for the entanglement entropies.
An excellent agreement is found between the lattice data and the 
corresponding analytic formulas in the continuum.

The manuscript is organized as follows. 
In \S\ref{sec:EHhom} we review the results for the homogenous systems
that will be extended to inhomogeneous systems, namely
some results of \cite{Cardy-Tonni16} for the entanglement hamiltonian 
and of \cite{cdt-17} for the contour function.
The rainbow chain is briefly introduced in \S\ref{sec:rainbow}.
In \S\ref{sec:EEinhom} we review the results of \cite{Dubail.17, Laguna.17}
about the entanglement entropies in inhomogeneous 1D free fermionic systems
for an interval adjacent to the boundary of a finite segment,
obtained through the twist field method.
Focussing on this configuration in the continuum limit of inhomogeneous 1D critical systems,
we construct the corresponding entanglement hamiltonian and entanglement spectrum in \S\ref{sec:EHI},
and in \S\ref{sec:contour} we provide the corresponding contour for the entanglement entropies. 
In \S\ref{sec:numerical} the analytic formulas derived in the previous sections
are specified for the rainbow model and checked against numerical data obtained for rainbow chains 
by adapting the methods of \cite{Fujita.17} and \cite{chen-vidal} to this inhomogeneous system. 
Conclusions are drawn in \S\ref{sec:conclusions}, where some open directions for future work are also discussed.


\section{Entanglement hamiltonian and contour in homogenous systems} 
\label{sec:EHhom}

In this section we review the results of \cite{Cardy-Tonni16} that are needed in the subsequent analysis.

Considering a subsystem $A$ made by an interval of finite length in a larger system which can have either infinite or finite size, 
the UV regularization of the path integral in the corresponding euclidean spacetime can be performed by 
removing infinitesimal circles of radius $\epsilon$ around the entangling points, namely the endpoints of $A$ and $B$
which separates $A$ from $B$.
The analysis of  \cite{Cardy-Tonni16} is valid in CFT and only for certain configurations where $A$ has either one or two entangling points. 
In the former case, the interval ends on the physical boundary of the whole system.

Once the UV regularization has been introduced, 
the configurations considered in \cite{Cardy-Tonni16} are such that 
one can construct a conformal map which sends the spacetime obtained
after the removal of the infinitesimal disks (parameterised by the complex coordinate $z$)
into an annulus described by another complex coordinate $w=f(z)$. 
This annulus is a rectangle whose width is $2\pi$ in 
the ${\rm Im} \,w$ direction and $W_{A}$ in the ${\rm Re} \, w$ direction,
where the annular structure is given by the 
identification ${\rm Im} \, w \sim {\rm Im} \, w + 2 \pi $.
A crucial role is played by the width $W_A$ of the annulus, which
is provided by the conformal map $f(z)$ as follows
\beq
W_A = \int_{A_\epsilon} f'(x) \, dx\,,
\label{3} 
\eeq
where $A_\epsilon$ corresponds to the interval left after the removal of the infinitesimal disks around the 
entangling points.
The width $W_A$ coincides with the difference between the values of
$f(x)$ at the endpoints of $A_\epsilon$ and it diverges logarithmically as $\epsilon \to 0$.
For instance, when $A=(u,v)$ is an interval of length $\ell=v-u$ on the infinite line, 
we have that $A_\epsilon =(u+\epsilon\, , v-\epsilon)$ 
and $W_A = 2\log(\ell/\epsilon)+ O(\epsilon)$.

The entanglement hamiltonian $K_A$ for the static configurations considered in \cite{Cardy-Tonni16} 
can be written in terms of the conformal map $f(z)$ as follows
\beq 
K_A = \int_A  \frac{T_{00}(x)}{f'(x)}  \,dx\,,
\label{4}
\eeq
where $T_{00} = T + \overline{T}$ is a specific component of the energy-momentum tensor
of the underlying CFT.
This leads to express the trace of the $n$-th power of the reduced density matrix as 
\beq
\Tr\, \rho^n_A \,=\, e^{- 2\pi n K_A}
\, =\,
 \frac{ \mathcal{Z}_{n \textrm{\tiny-annulus}} }{\mathcal{Z}_{\textrm{\tiny annulus}}^n}\,,
\label{29hom}
\eeq
where $\mathcal{Z}_{n \textrm{\tiny-annulus}}$ is 
the partition function of the underlying CFT where the worldsheet is an annulus 
similar to the one introduced above.
In particular, the $n$-annulus is a rectangle whose width is again $W_A$
 in the ${\rm Re} \, w$ direction, but the width in  the ${\rm Im} \,w$ direction
 is $2\pi n$.
 Thus, the identification ${\rm Im} \, w \sim {\rm Im} \, w + 2 \pi n$ is imposed for the 
 $n$-annulus.
 The boundary conditions are the same for both the annular partition functions
 occurring in (\ref{29hom}).

In the construction of \cite{Cardy-Tonni16}, 
the entanglement entropies can be written in terms of the width $W_{A}$ as follows
\beq
S^{(n)}_A = \frac{c}{12} \left( 1 + \frac{1}{n} \right) W_A + C_n+ o(1) \,,
\label{5}
\eeq
where $c$ is the central charge of the CFT and $C_n$ is a constant
depending on the boundary entropies introduced through the
regularisation procedure and therefore on the microscopic details of
the model. We remark that the well known logarithmic divergence of
$S^{(n)}_A$ as $\epsilon \to 0$ comes from $W_A$. The analysis of
\cite{Cardy-Tonni16} has access only to universal features; therefore the $n$
dependence of the non universal constant $C_n$ cannot be addressed.

In \cite{Cardy-Tonni16} also the eigenvalues $\lambda_{j}$ of the reduced density
matrix $\rho_A$ have been found. They are given by
\be
\label{lambda cft flat}
\lambda_{j}
=
\frac{q^{-c/24+\Delta_j}}{\mathcal{Z}_{\textrm{\tiny annulus}}}
=
\frac{q_{}^{-c/24+\Delta_j}}{\langle a | 0 \rangle  \langle 0 | b \rangle \, \tilde{q}_{}^{\,-c/24} }
\, \big( 1 + \dots \big) \,,
\ee
where the modular parameters of the flat annulus read
\beq
\label{mod par}
q
\equiv 
e^{-2\pi^2/W_{A} }  \,,
\qquad
\tilde{q}
\equiv 
e^{-2 W_{A} }  \,,
\eeq
and  $\Delta_j$ are the dimensions of the boundary operators
consistent with the boundary conditions at the edges of the annulus,
which are characterised by the boundary states $\ket|a>$ and $\ket|b>$. 
The dots in \eqref{lambda cft flat} correspond to subleading terms as $\epsilon \to 0$.  
In the intermediate step of (\ref{lambda cft flat}), the denominator
$\mathcal{Z}_{\textrm{\tiny annulus}}$ is the partition function of the underlying CFT on the
flat annulus.
From \eqref{lambda cft flat} and \eqref{mod par}
one easily obtains that
\bea
\label{lambda_j log}
- \log \lambda_{j}
&=&
- \left( \Delta_j - \frac{c}{24}\, \right) \log q
+ \log \mathcal{Z}_{\textrm{\tiny annulus}}
\\
\label{lambda_j expanded} 
&=&
\frac{c}{12}\, W_{A}  +\log \big( \bra<a| 0 \rangle  \bra<0| b \rangle \big)
+ \left( \Delta_j - \frac{c}{24}\right) \frac{2\pi^2}{W_{A}} 
+O(\epsilon^r )\,,
\eea
where $r>0$. 
Taking the limit $n \to 1$ of (\ref{5}), one finds that the entanglement entropy at leading order is given by $S_A = c \, W_A /6$.
Comparing this result with the expansion (\ref{lambda_j expanded}), it is straightforward to observe that
$- \log \lambda_{j} = S_A /2$ for any eigenvalue at leading order.

The largest eigenvalue corresponds to $\Delta_j = 0$, therefore we have
\be
- \log
\lambda_{\textrm{\tiny  max}}
\,=\,
\frac{c}{12}\, W_{A}  +\log \big( \bra<a| 0 \rangle  \bra<0| b \rangle \big)
- \frac{\pi^2 c}{12\,W_{A}} 
+O(\epsilon^r )\,,
\ee
which tells us that the leading divergence of  the maximum eigenvalue provides the central charge,
while the subleading one contains the boundary conditions.
The largest eigenvalue $\lambda_{\textrm{\tiny  max}}$ gives the single-copy entanglement and
the relation $- \log \lambda_{\textrm{\tiny  max}} = S_A /2$ in this case is well known \cite{single-copy-ent}.

We find it worth introducing the following gaps 
\beq
\log \lambda_{j} - \log \lambda_{k} 
=
\frac{2\pi^2}{W_{A} } \, (\Delta_k - \Delta_j)  \,,
\label{diff}
\eeq
where the r.h.s. has been obtained from \eqref{lambda_j log} and therefore 
it includes all orders corrections in  $\epsilon$. 
Considering the largest eigenvalue $\lambda_{j} = \lambda_{\textrm{\tiny  max}} $ 
and adopting the shorthand notation 
$\mathcal{E}_k \equiv \log \lambda_{\textrm{\tiny    max}} - \log \lambda_{k} $, we get
\beq
\label{eek} 
\mathcal{E}_k = \frac{2\pi^2   \Delta_k }{W_{A} } \,.
\eeq

We find it worth observing that, from (\ref{5}) and (\ref{eek}) we obtain
\beq
\label{consist rel1 flat}
S^{(n)}_A   \, \mathcal{E}_k  
\,=\,
\frac{ \pi^2}{6} \left( 1 + \frac{1}{n}\, \right) \Delta_k + \dots\,,
\eeq
where the dots denote infinitesimal terms as $\epsilon \to 0$.

The above discussion applies to all the static configurations such that
the euclidean spacetime obtained after the removal of the regularization disks
around the entangling point(s) can be mapped into an annulus \cite{Cardy-Tonni16}.

By plugging (\ref{3}) into (\ref{5}), it is straightforward to find that the entanglement entropies 
for the static configurations considered in \cite{Cardy-Tonni16} can be written as 
\beq
S_A^{(n)} = \int_{A_\epsilon}  s_A^{(n)}(x) \, dx\,,
\;\;\qquad\;\;
s_A^{(n)}(x)\geqslant 0\,,
\label{contour-cft-hom} 
\eeq
where
\be
\label{contour-func-cft-hom} 
s_A^{(n)}(x)  = 
\frac{c}{12} \left( 1 + \frac{1}{n} \right) f'(x)+ \frac{C_n}{\ell} \,.
\ee
In \cite{cdt-17} this expression has been proposed as the continuum limit of the contour function
for the entanglement entropies in CFTs for the static configurations discussed in \cite{Cardy-Tonni16}.

In this manuscript we focus on the case where $A$
is an interval of length $\ell$ adjacent to the boundary of the segment $(-L,L)$,
where the same boundary condition is imposed at the endpoints,
which is among the ones considered in \cite{Cardy-Tonni16}
(comparing to the notations, we have that $2L = L_{\textrm{\tiny there}}$).
We can set $A= (x_0, L)$ and only one entangling point occurs.

\begin{figure}[t!]
\vspace{.2cm}
\includegraphics[width=1.\textwidth]{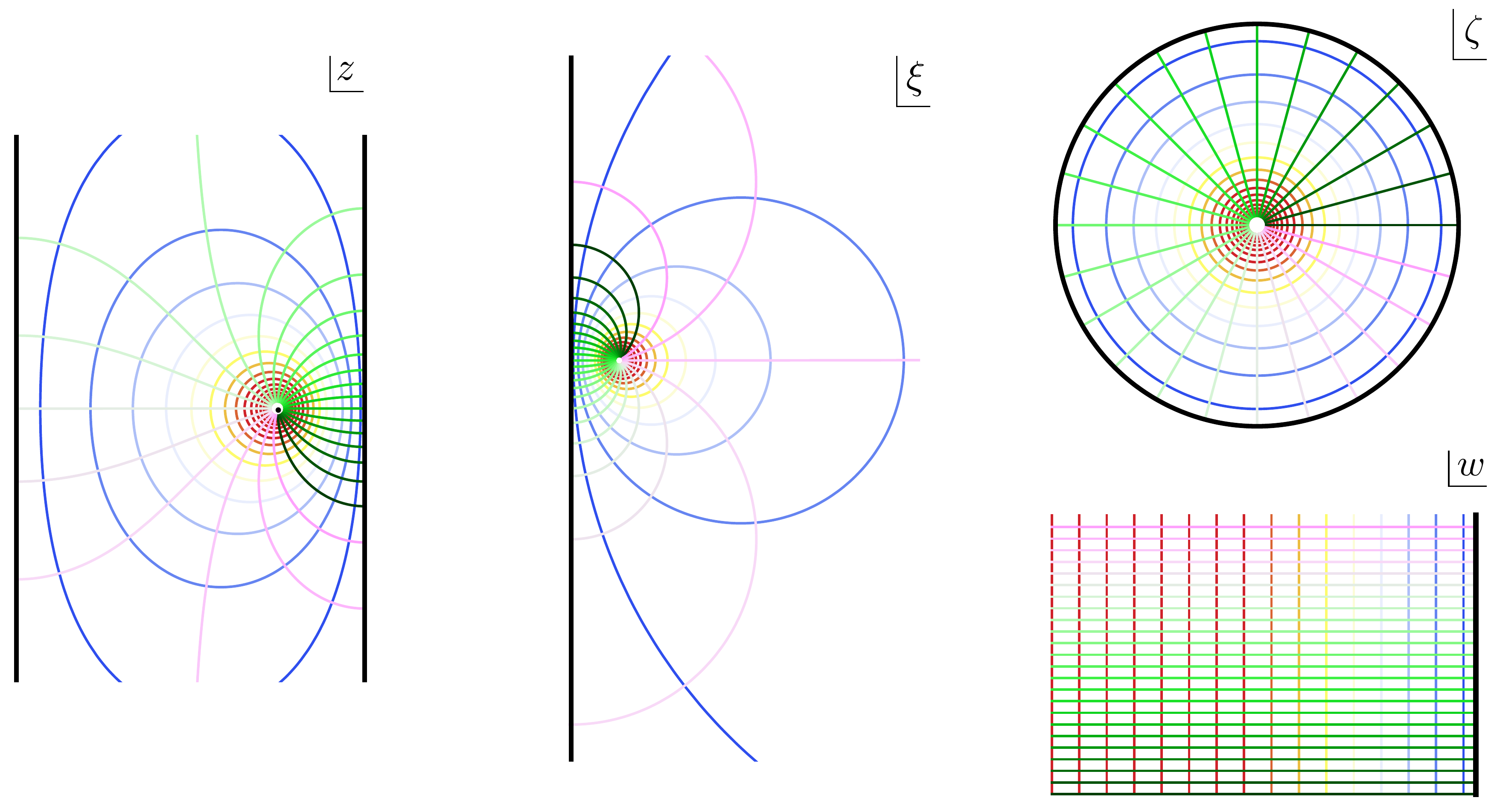}
\caption{
  Euclidean spacetimes describing the case of an interval $A
  =(x_0, L)$ of length $\ell = L- x_0$ in the segment $(-L,L)$. 
  In the  left panel a small disk $\D_{\epsilon}(x_0)$ of radius $\epsilon$ 
  centered at the entangling point $x_0$ has been removed from the strip 
  $\INT = \{(x,t_{\textrm{\tiny E}}) \in (-L,L) \times \Rmath \}$, where the same
  boundary condition is imposed on the boundaries (solid black lines).
  The sequence $z \to \xi \to \zeta \to w$ of conformal
  transformations allows to construct the conformal map \eqref{9},
  which sends the spacetime $\INT \setminus \D_{\epsilon}(x_0)$ 
  (left panel) into the annulus $\AN$ shown in the right bottom panel, 
  where the horizontal segment at $\textrm{Im} \,w = 0$ and the one at 
  $\textrm{Im} \,w = 2\pi$ delimiting the spacetime in the vertical direction 
  must be identified. 
  The width $W_A$ of the annulus $\AN$ is given by \eqref{10}. 
  }
   \label{fig:map}
\end{figure}

For this configuration we have to consider the euclidean spacetime given by the strip 
$\INT = \{( x,t_{\textrm{\tiny E}}) \in (-L,L) \times \Rmath \}$, which will be
parameterised by the complex coordinate $z = x+ \iu t_{\textrm{\tiny E}}$ in the following. 
The UV regularization employed in \cite{H94, Cardy-Tonni16}
requires to remove the infinitesimal disk
$\D_{\epsilon}(x_0) = \{ |z-x_0| \leqslant \epsilon \}$
of radius $\epsilon$ centered at the entangling point $z=x_0$.
The resulting spacetime $\INT \setminus \D_{\epsilon}(x_0)$ is shown in the 
left panel of Fig.\,\ref{fig:map}.
This regularization procedure naturally leads to introduce the regularised intervals
$A_\epsilon \equiv (x_0 + \epsilon, L)$ and $B_\epsilon \equiv (-L, x_0 - \epsilon)$.

The conformal map $w=f(z)$ which sends the holed strip 
$\INT \setminus \D_{\epsilon}(x_0)$ into the annulus  
(right bottom panel in Fig.\,\ref{fig:map}) can be constructed through
the following intermediate steps. 
First one maps the strip $\INT$ into the right half plane 
(middle panel in Fig.\,\ref{fig:map}) through 
the conformal transformation $z \to \xi = e^{\iu \pi z/(2L)}$.
The entangling point $x_0$ is mapped into $\xi_0 \equiv \xi(x_0)$.
Then, the resulting spacetime can be mapped into 
the circular crown $\D_1 \setminus \D_R$ (right top panel in Fig.\,\ref{fig:map}),
being $\D_1 = \{ \zeta , |\zeta| \leqslant 1 \}$ the unit disk
and $\D_R = \{\zeta, |\zeta| \leqslant R \}$ the infinitesimal disk of radius $R$ centered in the origin.
The conformal map $\xi \to \zeta (\xi)$ implementing this transformation must send $\xi_0$ into the origin.
It reads \cite{T13, Cardy-Tonni16}
\beq
\zeta(z)  
= -\,\iu\, \frac{(1+ \bar{\xi}_0)(\xi-\xi_0)}{(1+ \xi_0)(\xi + \bar{\xi}_0)}
= \frac{ \sin( \pi ( z - x_0)/(4 L))}{ \cos( \pi ( z + x_0)/(4 L))}  \,.
\label{7}
\eeq 
The boundaries of $\INT$, which have $x = -L$ and $x=L$, are sent into
the boundary of $\D_1$, which is the circumference $|\zeta| = 1$.  
The disk $\D_{\epsilon}(x_0) \subset \INT$ is mapped into the disk $\D_R \subset \D_1$ centered in the origin 
whose radius is infinitesimal as $\epsilon \to 0$;
indeed
\beq
R \simeq  \frac{\pi \epsilon/(4 L)}{ \cos( \pi x_0/(2 L))} = \frac{\pi \epsilon/( 4 L)}{ \sin( \pi \ell/(2 L))} \,,
\label{8}
\eeq 
being $\ell = L - x_0$ the length of the interval $A$.  The intervals
$A$ and $B$ are mapped into the segments $(R, 1)$ and $(-1, -R)$
respectively.

Finally, the conformal transformation  $w= f(z)$  is given by \cite{Cardy-Tonni16} 
\beq
w= f(z) = \log \big( \zeta(z) \big) 
= 
\log  \left( \frac{ \sin( \pi ( z - x_0)/(4 L))}{ \cos( \pi ( z + x_0)/(4 L))} \right)  ,
\label{9}
\eeq 
which sends  $\INT \setminus \D_{\epsilon}(x_0)$ into the
annulus $\AN= (\log R, 0) \times [0,2\pi)$, where ${\rm Im} \, w=0$ is
  identified with ${\rm Im} \, w=2 \pi$ (see the right bottom panel in
  Fig.\,\ref{fig:map}).  
  The regularised interval $A_\epsilon$ is mapped into the  interval $[\log R, 0]$ on the negative real axis,
  while the block $B_\epsilon$ lies in the center of the annulus having ${\rm Im } \, w= \pi$.

 The general expressions discussed above for the entanglement hamiltonian, the entanglement spectrum
 and the contour for the entanglement entropies can be specified to this configuration by employing the map (\ref{9}).
  In this case the boundary states $\ket|a>$ and $\ket|b>$ encode the two (maybe different) boundary conditions that must be specified:
along the vertical lines, which correspond to the physical boundary of the segment,
and along the boundary of the infinitesimal disk $ \D_{\epsilon}(x_0)$ centered at the entangling point.

The weight function to employ in the expression (\ref{4}) of the entanglement hamiltonian
for this configuration can be easily obtained from (\ref{9}).  It reads 
\beq
\frac{1}{f'(x)} 
 \,=\,
\frac{2L}{\pi}\;
\frac{\sin(\pi x/(2L) ) - \sin(\pi x_0/(2L) )}{\cos(\pi x_0/(2L) )}\,.
\label{weight function homo}
\eeq
From this formula it is straightforward to notice that $f'(x)>0$ for $x\in A$.

The expansion of this weight function as $x \to x_0$ is given by
\be
\frac{1}{f'(x)} = (x-x_0) -  \frac{\pi \tan(\pi x_0 / (2L))}{4L} \,(x-x_0)^2 + O\big((x-x_0)^3 \big)\,.
\label{weight exp}
\ee
The leading term corresponds to the expected behavior dictated by the
Bisognano-Wichmann theorem for a semi-infinite line \cite{bw}. 
As for the subleading $O((x-x_0)^2 )$ term, its sign is opposite to the sign of $x_0$;
therefore when $x \simeq x_0^+$ we have that the curve (\ref{weight exp}) is above 
the one provided by the Bisognano-Wichmann result for $x_0 < 0$,
while it stays below for $x_0 >0$.

From (\ref{9}), we have that the width of the $\AN$ (right bottom panel of Fig.\,\ref{fig:map}) is \cite{Cardy-Tonni16}
\beq
W_A  = f(L) - f(x_0 +  \epsilon) = - \log R = \log \left[ \frac{ 4 L}{ \pi \epsilon} \sin \left( \frac{ \pi \ell}{  2 L} \right)  \right]  + O(\epsilon)\,.
\label{10}
\eeq 
By employing this result into (\ref{5}) we obtain  the entanglement entropies  
\beq
S^{(n)}_A = \frac{c}{12} \left( 1 + \frac{1}{n} \right)  \log \left[  \frac{ 4 L}{ \pi \epsilon} \sin \left( \frac{ \pi \ell}{  2 L} \right)  \right]     + C_n  + o(1)\,.
\label{11}
\eeq
For this configuration the gaps (\ref{eek}) become
\beq
{\cal E}_k =   
\frac{ 2 \pi^2  \Delta_k
}{ 
\log \big[  \tfrac{4 L}{ \pi \epsilon } \,\sin( \tfrac{ \pi \ell}{  2 L} )  \big]  } \,.
\label{12}
\eeq
As consistency check, notice that for $A$ given by half system (i.e. $\ell = L$)
one recovers the well known scaling dependence $1/\log L$ 
\cite{ent-ham-latt, TP88}. 

Finally, let us write explicitly also the contour function for the entanglement entropies 
corresponding to this configuration, which can be easily obtained from
(\ref{contour-func-cft-hom}) and (\ref{weight function homo}). The result is
\be
\label{contour-func-strip} 
s_A^{(n)}(x)  = 
\frac{c}{12} \left( 1 + \frac{1}{n} \right) 
\frac{\pi}{2L}\;
\frac{\cos(\pi x_0/(2L) )}{\sin(\pi x/(2L) ) - \sin(\pi x_0/(2L) )} + \frac{C_n}{\ell} \,.
\ee
From (\ref{weight exp}), we have that $s_A^{(n)}(x) = \tfrac{c}{12}(1+\tfrac{1}{n})/(x - x_0) + \dots$ as $x \to x_0^+$.
Instead, it is straightforward to notice $s_A^{(n)}(L)$ is finite. 
We remark that the product $L\, s_A^{(n)}(x)$ is a function of $x/L$ parameterised by $n$ and $x_0 /L$.
In \cite{cdt-17} the profile (\ref{contour-func-strip}) has been considered as the continuum limit 
of a contour for the entanglement entropies in the massless harmonic chain with Dirichlet boundary condition
imposed at both the endpoints of the segment.


\section{The rainbow model} 
\label{sec:rainbow}

A  relevant family of fermionic models in one spatial dimension is characterised by the following hamiltonian
\beq
H=- \frac{1}{2} \sum_i J_i\;c^\dagger_i c_{i+1} + \mbox{h.c.}\,,
\label{eq:J_ham}
\eeq
where $c^\dagger_i$ creates a fermionic particle on the $i$-th site 
and $J_i >0$ are inhomogeneous hopping parameters. 
In the case of strong inhomogeneity, the ground state of the hamiltonian (\ref{eq:J_ham}) 
is  a valence bond state that can be studied via the strong disorder renormalization group 
of Dasgupta and Ma \cite{Dasgupta.80,Ramirez.14b}. 
The entanglement entropy of a given block in this regime can be obtained just by counting the number of bonds cut 
when the partition between the subsystem and its complement is introduced \cite{rm-04}.

If the hopping parameters $J_i$ are carefully engineered, this valence bond state 
becomes a ground state whose entanglement entropy satisfies a volume law. 
This phenomenon occurs for the rainbow chain 
\cite{Vitagliano.10,Ramirez.14b,Ramirez.15,Laguna.16,Laguna.17}
which is defined on a lattice made by $2L$ sites and whose hamiltonian belongs to the class defined by (\ref{eq:J_ham}).
In particular, the hamiltonian of the rainbow chain is
\beq
  H =  
  -\frac{J}{2} \,c_{\frac{1}{2}}^\dagger c_{-\frac{1}{2}} 
  - \frac{J}{2}  \,\sum_{m=\frac{1}{2}}^{L -\frac{3}{2}} e^{-hm}
\left[c^\dagger_m c_{m+1} +c^\dagger_{-m} c_{-(m+1)}\right] + \textrm{h.c.}\,,
  \label{eq:rainbow_ham}
\eeq
where the sites indices have been shifted to half-integers for simplicity.
The parameter $J>0$ sets the energy scale and $h\geqslant 0$ characterizes
the inhomogeneity of the hopping amplitudes. 
After a Jordan-Wigner transformation, the hamiltonian \eqref{eq:rainbow_ham} 
becomes the hamiltonian of an inhomogeneous spin-1/2 XX chain \cite{Vitagliano.10}.
The case $h=0$ corresponds to a standard uniform spinless free fermion (tight binding model)
with open boundaries, whose low energy properties are described by a boundary CFT with central charge $c=1$ 
and Luttinger parameter $K=1$.

\begin{figure}[t!]
  \centering
  \vspace{.5cm}
  \includegraphics[width=.8\textwidth]{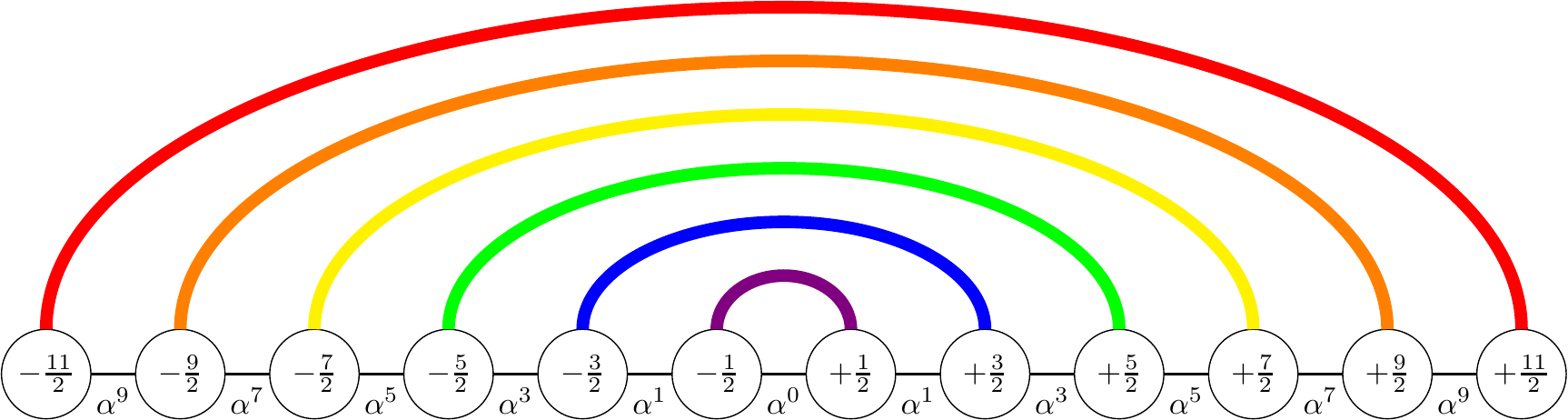}
    \vspace{.2cm}
  \caption{
    Rainbow state as valence bond state, where the bonds above the central
    link connect sites in symmetric position with respect to the center of the chain. 
    The entanglement entropy between the left and the
    right halves of the chain is $L \log 2$, being $2L$ the total number of sites. 
    In this figure  $\alpha\equiv e^{h/2}$.}
  \label{fig:RB_scheme}
\end{figure}

 In \cite{Vitagliano.10,Ramirez.14b,Ramirez.15} the hamiltonian \eqref{eq:rainbow_ham} 
has been studied in the limits of strong and in the weak inhomogeneity, 
which correspond to $h \gg 1$ and $h \ll 1$ respectively. 

In the strong inhomogeneity limit $h \gg 1$, the application of the
strong disorder renormalization group  algorithm
leads to a valence bond state built from singlets between the sites $m$ and $-m$, 
for $m= \frac{1}{2}, \dots , L- \frac{1}{2}$ 
(in the XX version of the hamiltonian). 
Thus, the ground state resembles a rainbow like the one schematically represented in Fig.\,\ref{fig:RB_scheme}, 
where the colours of the bonds are associated to the energy required to break them, which is proportional
to $e^{-2mh}$. 
The entanglement entropy between
the left and right halves on this state is $S_A = L \log 2$, which
corresponds to the highest value reachable for a subsystem containing
$L$ qubits. 
In the limit $h \to \infty$ the rainbow state becomes the
exact ground state of the hamiltonian \eqref{eq:rainbow_ham}, 
but several of its properties persist for all values of $h$. 
In particular, the entanglement entropy grows linearly with $L$ with a slope which depends on $h$.

In the limit $h \ll 1$ of weak inhomogeneity, the low energy physics
of the hamiltonian \eqref{eq:rainbow_ham} is described by the
following hamiltonian of two chiral fermions $\psi_L$ and $\psi_R$
\cite{Ramirez.15}
\be
  H  \simeq    \iu   J a \int_{-a  L}^{a L} dx \, e^{-\frac{h |x|}{a}} \left[
  \psi^\dagger_R \, \partial_x \psi_R -\psi^\dagger_L \, \partial_x \psi_L
   -  \frac{h}{2 a} {\rm sign}{(x)} 
   \Big(\psi^\dagger_R \psi_R  -\psi^\dagger_L \psi_L \Big) \right],
    \label{ham rb2}
\ee
where $a$ is the lattice spacing, $x = m a$ is the position and the
fields $\psi_{L,R}(x)$ are the slow varying modes of the fermion
operator $c_m$ expanded around the Fermi momenta $\pm k_F$ (where $k_F
= \pi/(2a)$ at half filling), namely
\beq
  \frac{c_n}{\sqrt{a}} \,\simeq \,e^{\iu k_F x} \,\psi_L(x) +e^{-\iu  k_F x} \,\psi_R(x) \,.
  \label{rainbow2}
\eeq
The continuum limit corresponds to the regime where $a \rightarrow 0$,
$h \rightarrow 0$ and $L \rightarrow \infty$, with $h/a$ and $aL$ kept
constant. This implies that the product $\lambda \equiv h L$
is independent on the lattice spacing.
Several quantities depend on this combination. 
We find it convenient to rename $h/a \rightarrow h$ and
$aL \rightarrow L$. Thus, $h$ and $L$ acquire the dimensions
respectively of an inverse length and a length. 
In  \eqref{ham rb2} the fields $\psi_{L}$ and $\psi_{R}$ are decoupled in the bulk but their coupling comes from the boundary conditions
\cite{Ramirez.15}
\beq
  \label{rainbow4}
  \psi_R (\pm L) = \mp \,\iu \,  \psi_L (\pm  L)\,,
\eeq
which are imposed at the endpoints of the segment. 
%


\section{Entanglement entropies for inhomogeneous fermionic systems}
\label{sec:EEinhom}

The powerful techniques of CFT in curved backgrounds  have been successfully applied 
to study entanglement entropies of some configurations in some inhomogeneous fermionic systems
such as the gas of free fermions trapped by the harmonic potential \cite{Dubail.17}
and the rainbow model \cite{Ramirez.14b,Ramirez.15,Laguna.17}. 
The conclusion of these works is that the long
distance behavior of these models can be described by a massless Dirac
fermion in a curved spacetime whose metric encodes the inhomogeneity
parameters. 
The analytic expressions obtained for the entanglement entropies agree with
high accuracy with the corresponding numerical results obtained
through standard techniques.

The field theory we need to consider is a free massless Dirac fermion
on the strip $\INT$. Its two dimensional euclidean action reads
\beq
S = \int_\INT \frac{ d z d\bar{z}}{2 \pi} \; e^{ \sigma(x)} \left[ \psi^\dagger_R  \stackrel{\leftrightarrow}{\partial_{\bar z}}  \psi_R  + 
\psi^\dagger_L  \stackrel{\leftrightarrow}{\partial_z}  \psi_L \right] ,
\label{13}
\eeq 
where the background metric is given by 
\beq
ds^2 = e^{2 \sigma(x)} dz d \bar{z} \,,
\label{14}
\eeq 
which is Weyl equivalent to the flat metric, with the Weyl factor $e^{\sigma(x)}$ depending only on the spatial variable. 
We remark that in the analysis of \cite{Dubail.17,Ramirez.14b} the worldsheet is the whole strip $\INT$,
while the worldsheet introduced in \S\ref{sec:EHhom} is the strip 
where infinitesimal disks centered at the entangling points have been removed.

Since the Weyl factor does not depend on the Euclidean time $t\equiv t_{\textrm{\tiny E}}$,
the metric (\ref{14}) has a timelike Killing vector which guarantees 
that the corresponding  Dirac hamiltonian derived from \eqref{13} is conserved. 
The complex coordinate $z$ in \eqref{14} is given in terms of coordinates $(x,t)$ as follows
\beq
z =  \tilde{x} + \iu\,t\,,  \qquad \tilde{x} = \int_0^{x} e^{ - \sigma(y)}\, dy  \,,
\label{15}
\eeq
where $\tilde{x}$ is such  that $\tilde{x}'(x) = e^{-\sigma(x)}$. 
The scalar curvature of the metric (\ref{14}) is given by
\beq
{\cal R}
= -\, 2 \,e^{ - 2  \sigma(\tilde{x} )} \, \partial_{\tilde{x}}^2 \sigma(\tilde{x}) 
= 
- \,2 \big[ \big(\partial_x \sigma(x)\big)^2 + \partial_x^2 \sigma(x) \big]  \,,
\label{15b}
\eeq
where (\ref{15}) has been employed in the last step.
We consider  background metrics that are symmetric with respect to the center $x=0$ of the segment, namely  $\sigma(-x) = \sigma(x)$.
In these cases we have
\beq
 \tilde{x} \in (- \tilde{L},  \tilde{L} ) \,,
  \;\;\qquad \;\;
  \tilde{L} \equiv \int_0^L e^{ - \sigma(y)}\, dy   \,.
\label{16}
\eeq

For the Fermi gas trapped in a harmonic potential, the Weyl factor is
the local Fermi velocity $e^{\sigma(x)}=v_F(x)$, which is given by
\cite{Dubail.17}
\beq
e^{\sigma(x)} = \sqrt{L^2- x^2}  
\qquad \Longrightarrow \qquad 
\tilde{x} = \arcsin \frac{x}{L}\,,
\qquad \tilde{L} = \frac{\pi}{2}   \,.
\label{17}
\eeq 
The scalar curvature of this background is ${\cal R} = 2/ [ L^2
  (\cos\tilde{x})^4]$.  For an arbitrary external potential $V(x)$,
one has $e^{ \sigma(x)} = v_F(x) = \sqrt{ 2 ( \mu - V(x))}$, where
$\mu$ is the number of particles in the gas.

In the rainbow model, the Weyl factor encodes the local hopping
amplitudes decreasing exponentially from the center towards the
edges of the segment \cite{Ramirez.14b}. It reads
\beq
e^{\sigma(x)} = e^{-h|x|} 
\qquad \Longrightarrow \qquad 
\tilde{x} = {\rm sign} (x) \frac{ e^{ h |x|}  -1}{h}\,, \qquad \tilde{L} =   \frac{ e^{ h L } -1}{h}   \,,
\label{18}
\eeq
where $h \geqslant 0$. 
The scalar curvature of this background is ${\cal R}  = - 2 h^2 + 4 h \delta(\tilde{x})$, namely it is constant and negative everywhere except for a singularity at the origin.
These inhomogeneous models become uniform by taking a specific limit.

The entanglement entropies for an interval $A=(x_0, L)$ 
adjacent to the boundary of the segment $(-L, L)$ 
in these inhomogeneous fermionic systems  have been computed 
by employing the twist field method \cite{Dubail.17}. 
The result is 
\beq
{\rm Tr} \, \rho_A^n = \eta^{\Delta_n} \langle {\cal T}_n(x_0, 0) \rangle_{\textrm{\tiny curved}}    \,,
\label{19}
\eeq
where ${\cal T}_n(x_0,0)$ is a twist field  at the entangling point $(x,t_{\textrm{\tiny E}})=(x_0, 0)$, 
whose dimension $\Delta_n$ is given by  (with $c=1$ for a free fermion)
\beq 
\Delta_n = \frac{1}{12} \left(n - \frac{1}{n} \right)  ,
\label{20}
\eeq 
and $\eta$ is a UV cutoff, whose relation with the cutoff $\epsilon$ 
introduced in \S\ref{sec:EHhom} is discussed below. 

The expectation value \eqref{19} is computed with the action
\eqref{13} that contains the Weyl factor $e^{\sigma}$.  
A Weyl rescaling transforms the metric \eqref{14} into the flat metric $d z d
\bar{z}$. 
Under this transformation the twist operator transforms as
\beq
{\cal T}_n(x_0, 0) 
\,\rightarrow \,
\big(\tilde{x}'(x_0)\big)^{ \Delta_n}  \, {\cal T}_n(\tilde{x}_0, 0)
 \,= \,
 e^{- \Delta_n \sigma(x_0)} \, {\cal T}_n(\tilde{x}_0, 0)   \,.
\label{21} 
\eeq  
Thus \eqref{19} becomes
\beq
\Tr \, \rho_A^n =
\eta^{\Delta_n}  e^{- \Delta_n \sigma(x_0)}
\langle {\cal T}_n( \tilde{x}_0, 0) \rangle_{\textrm{\tiny flat}}\,,
\label{22}
\eeq 
where $\tilde{x}_0 \equiv \tilde{x}(x_0)$. 
The correlator $\langle {\cal T}_n(\tilde{x}_0, 0) \rangle_{\textrm{\tiny flat}}$ is
computed on the strip $(-\tilde{L}, \tilde{L}) \times \Rmath$ with
flat background, which can be mapped into the flat upper half plane
(UHP) through the transformation $z \rightarrow g(z) = e^{ \iu \pi (
  \tilde{L}-z)/( 2 \tilde{L})}$;  therefore we have
\beq
\Tr \, \rho_A^n = 
\eta^{\Delta_n}  e^{- \Delta_n \sigma(x_0)}  
\left|\frac{d g}{d \tilde{x}}   \right|^{\Delta_n}_{\tilde{x} = \tilde{x}_0}  
 \langle {\cal T}_n( g(\tilde{x}_0),   \bar{g}(\tilde{x}_0) ) \rangle_{\textrm{\tiny uhp}}  \,.
\label{23}
\eeq
By employing the following correlator in the upper half plane
\beq
\langle {\cal T}_n( g(\tilde{x}),\bar{g}(\tilde{x}) ) \rangle_{\textrm{\tiny uhp}}  
= ({\rm Im} \, g(\tilde{x}) )^{ - \Delta_n}
= \left( \sin \frac{ \pi ( \tilde{L} - \tilde{x} )}{   2 \tilde{L}} \right)^{ - \Delta_n} ,
\label{24}
\eeq
one obtains
\beq
\Tr \, \rho_A^n =   \left[   \,  e^{ \sigma(x_0)} \, \frac{ 2 \tilde{L} }{ \pi \eta} \,
 \sin \frac{ \pi ( \tilde{L} - \tilde{x}_0 )}{   2 \tilde{L}} \, \right]^{ - \Delta_n}   .
\label{25}
\eeq 
This gives the following expression for the entanglement entropies
\beq
S^{(n)}_A =    
\frac{1}{12} \left( 1 + \frac{1}{n} \right)  
\log \left[ \, e^{ \sigma(x_0)} \,
\frac{ 2 \tilde{L} }{ \pi \eta} \,  \sin \frac{ \pi ( \tilde{L} - \tilde{x}_0 )}{   2 \tilde{L}} \,\right]  ,
\label{26} 
\eeq 
which coincides with  the formula of the entanglement
entropies of an interval $(\tilde{x}_0 , \tilde{L})$ in the segment
$(- \tilde{L}, \tilde{L})$ for the homogenous models, except for the
additional constant term due to the factor $e^{ \sigma(x_0)} $ inside
the argument of the logarithm.

As for the Fermi gas  trapped in a harmonic potential, 
where the density depends on the position,
in \cite{Dubail.17}  it has been explained that a further spatial dependence 
must be introduced in the UV cutoff as $\eta \to \eta(x) = \eta_0 / k_{\textrm{\tiny F}}(x)$,
being $k_{\textrm{\tiny F}}(x) = \sqrt{L^2 - x_0^2}$ the Fermi momentum.
By employing this important observation and  \eqref{17} in \eqref{26}, one finds \cite{Dubail.17} 
\beq
S^{(n)}_A =     
\frac{1}{12} \left( 1 + \frac{1}{n} \right)  
 \log \left[  
\frac{ L ^2}{\eta_0} \left( 1 - \frac{ x^2_0}{L^2} \right)^{3/2}\, \right] .
\label{27}
\eeq 
In the rainbow model, where $k_{\textrm{\tiny F}} = \pi/2$ and the inhomogeneity is due to the spatial dependence of the couplings, 
there is no need to introduce a spatial dependence for the UV cutoff $\eta$;
therefore, by using \eqref{18} in \eqref{26}, 
one obtains \cite{Laguna.17} 
\beq
S^{(n)}_A =     
\frac{1}{12} \left( 1 + \frac{1}{n} \right)  
\log \left[  \, e^{- h |x_0| } 
\frac{ 2 ( e^{h L} -1)  }{ \pi  h \, \eta  } \,  \cos \left( \frac{\pi}{2}  \frac{e^{h|x_0|}-1}{e^{hL}-1}  \right)   \right]  .
\label{28}
\eeq

In \cite{Dubail.17} and \cite{Laguna.17} the entanglement entropies of free fermions trapped by the harmonic potential
and of the rainbow chain respectively have been studied also numerically through standard techniques for free fermions on the lattice.
Complete agreement has been obtained between with the analytic expressions reported in \eqref{27} and \eqref{28}.


\section{Entanglement hamiltonian for inhomogeneous systems}
\label{sec:EHI}

In this section we employ the results of \cite{Cardy-Tonni16} reviewed in \S\ref{sec:EHhom} 
to study the entanglement hamiltonian and the corresponding entanglement spectrum for 
the interval $A=(x_0,L)$ in the segment $(-L, L)$
in  the  inhomogeneous systems described in \S\ref{sec:rainbow} and \S\ref{sec:EEinhom}.

The analysis of \cite{Cardy-Tonni16} has been carried out in flat spacetimes, where $ds^2 = dz d \bar{z}$.
In curved spacetimes with background metric $ds^2 = e^{2 \sigma} dz d \bar{z}$, 
the regularisation procedure and the conformal mappings described in \S\ref{sec:EHhom}  
can be repeated, finding that
\beq
\Tr\, \rho^n_A \,=\, e^{- 2\pi n K_A}
\, =\,
 \frac{ \mathcal{Z}_{n \textrm{\tiny-annulus, curved}} }{\mathcal{Z}_{\textrm{\tiny annulus, curved}}^n}\,,
\label{29}
\eeq
which is different from (\ref{29hom}) because the curved background metric occurs in the
annular partition functions. 

In the appendix\;\ref{app:liouville} we employ the characteristic property of Liouville theory 
\cite{polyakov-81, alvarez-82} to argue  that 
$\mathcal{Z}_{\textrm{\tiny annulus, curved}} = \Omega\, \mathcal{Z}_{\textrm{\tiny annulus, flat}}$ 
and $\mathcal{Z}_{n \textrm{\tiny-annulus, curved}} =  \Omega^{n} \mathcal{Z}_{n \textrm{\tiny-annulus, flat}}$,
where $\Omega$ is a positive factor. 
Focussing on the inhomogeneous systems whose background is described by (\ref{14}),
this observation implies that $\Tr \rho_A^n$ is given by the same  expression found for the flat geometry
in terms of  the coordinates $(\tilde{x}, t)$ (see (\ref{15})).
This result can be obtained if the same construction holds also for the entanglement hamiltonian. 
In particular, from \eqref{4}
we can write the entanglement hamiltonian $K_A$ in the inhomogeneous systems
characterised by (\ref{14}) as follows 
\beq
K_A = \int_{\tilde{A}} \frac{T_{00}(\tilde{x}) }{\tilde{f}'( \tilde{x})} \; d \tilde{x}\,,
\label{30}
\eeq
where the integration domain is $\tilde{A} \equiv (\tilde{x}_0 , \tilde{L} )$, which 
can be found from (\ref{15}) and (\ref{16}).
In (\ref{30}) we have denoted by $\tilde{f}(x)$ the function obtained from $f(x)$ by replacing all the parameters 
with the corresponding tilded ones. 
By employing the transformation law $T_{00}(\tilde{x}) = \tilde{x}'(x)^{-2} \, T_{00}(x)$ 
for the component of the energy-momentum tensor
(the schwarzian derivative term can be neglected in this analysis, as argued in \cite{Cardy-Tonni16}), 
the entanglement hamiltonian \eqref{30} can be expressed in terms of the original  spatial variable $x$ as follows
\beq
K_A = \int_A  \beta_A(x) \,  T_{00}(x) \, dx\,,
 \label{30a}
\eeq
where $A =(x_0, L) $ and the weight function $\beta_A(x)$ is given by 
\beq
\beta_A(x) \,=\, \frac{1}{ \tilde{x}'(x) \, \tilde{f}'(\tilde{x}(x)) }\,,
\label{30b}
\eeq 
which can be also written as $1/\beta_A(x) = \tfrac{d \tilde{f}(\tilde{x}(x))}{dx}$.
Since from (\ref{15}) we have $\tilde{x}'(x) = e^{-\sigma(x)}$, the positivity of $\beta_A(x) $
comes from the fact that $f'(x)$ is positive (see (\ref{weight function homo})).

The expansion of $f'(x)$ as $x \rightarrow x_0^+$ for the functions that we are considering is given by $f'(x) = (x-x_0)^{-1} + f_0 + O(x-x_0)$.
As for (\ref{30b}) in this limit, we  find that 
\be
\label{beta expand x0}
\beta_A(x) 
\,=\,
(x-x_0) 
- \left(
\tilde{x}'(x_0) \, \tilde{f}_0 +\frac{\tilde{x}''(x_0) }{2\, \tilde{x}'(x_0) }
\right)
(x-x_0)^2 + \dots\,,
\ee
where $\tilde{f}_0$ is the coefficient occurring in the expansion $\tilde{f}'(x) = (x-x_0)^{-1} + \tilde{f}_0 + O(x-x_0)$ as $x \to x_0^+$
and the dots denote subleading terms.
The leading term in (\ref{beta expand x0}) corresponds to the characteristic behaviour described by the Bisognano-Wichmann theorem
and it does not depend on the features of the underlying model, occurring instead in the subleading term. 
By using that $\tilde{x}'(x) = e^{-\sigma(x)}$, which implies $\tilde{x}''(x)/ \tilde{x}'(x)= -\, \sigma'(x)$,
we obtain that the coefficient of $(x-x_0)^2$ in (\ref{beta expand x0}) simplifies to $ \sigma'(x_0)/2 - e^{-\sigma(x_0)} \tilde{f}_0$.

For the configuration given by the interval $A=(x_0, L)$ in the segment $(-L, L)$,
the conformal map to consider is (\ref{9}).
The weight function (\ref{30b}) for the entanglement hamiltonian can be written explicitly by employing (\ref{weight function homo}), finding
\beq
\beta_A(x) \,=\,
\frac{2\tilde{L}}{\pi}\;
\frac{\sin(\pi \tilde{x}(x)/(2\tilde{L}) )
  - \sin(\pi \tilde{x}_0/(2\tilde{L}) )}
     {\cos(\pi \tilde{x}_0/(2\tilde{L}) ) \, \tilde{x}'(x)}\,,
\label{ent temp xtilde}
\eeq
where $\tilde{x}_0 = \tilde{x}(x_0)$.
We remark that $\beta_A(x) $ does not contain the UV cutoff.

The expansion (\ref{ent temp xtilde}) as $x\to x_0^+$ is (\ref{beta expand x0}), 
where the coefficient of $(x-x_0)^2$ can be written explicitly by using (\ref{18}) and that 
$\tilde{f}_0 = \tfrac{\pi}{4\tilde{L}} \tan(\tfrac{\pi \tilde{x}_0}{2\tilde{L}})$ (see (\ref{weight exp})).

The above discussion naturally leads to consider
the width of the flat annulus in terms of the tilded quantities.
From \eqref{10}, it reads
\beq
\widetilde{W}_A   =  
\log \bigg[ \frac{ 4  \tilde{L}}{ \pi {\tilde{\epsilon}}}   \sin \bigg( \frac{ \pi \tilde{\ell}}{  2  \tilde{L}} \bigg)  \bigg] + O(\tilde\epsilon) 
\,,
\;\qquad \;
\tilde{\ell} = \tilde{L} - \tilde{x}_0 \,.
\label{33}
\eeq 

The UV cutoff $\tilde{\epsilon}$ in this expression plays a crucial role in our analysis.
It can be written in terms of the physical UV cutoff $\epsilon$ by using that the endpoints
of the interval $\tilde{A}_{\tilde{\epsilon}}= (\tilde{x}_0 + \tilde{\epsilon}, \tilde{L})$ are related 
to the corresponding endpoints of $A_\epsilon= (x_0 + \epsilon, L)$.
In particular, from the first endpoints of these two intervals, we have
\beq
\tilde{x}(x_0 + \epsilon) = \tilde{x}_0 + \tilde{\epsilon}  + O(\epsilon^2) 
\qquad \Longrightarrow \qquad
\tilde{\epsilon} =  \, e^{- \sigma(x_0)} \, \epsilon\,,
\label{31}
\eeq
where $\tilde{x}'(x_0) = e^{ - \sigma(x_0)}$ has been employed. 
We remark that the inhomogeneity in some models (e.g. the trapping potentials in \cite{Dubail.17})
requires to introduce a UV cutoff $\epsilon$ which depends on $x_0$.
This dependence cannot be captured through our CFT analysis.

The entanglement entropies for the interval $A=(x_0, L)$ in the segment $(-L,L)$ in the inhomogeneous models
can be easily written from (\ref{5}), \eqref{33}  and \eqref{31}, finding 
\beq
S^{(n)}_A   = 
\frac{c}{12} \left( 1 + \frac{1}{n} \right)  
\widetilde{W}_A
\,=\, 
\frac{c}{12} \left( 1 + \frac{1}{n} \right)    
 \log \bigg[  e^{ \sigma(x_0)} \; \frac{ 4  \tilde{L}}{ \pi {\epsilon}}   \sin \bigg( \frac{ \pi \tilde{\ell}}{  2  \tilde{L}} \bigg)  \bigg]\,,
\label{34}
\eeq
up to a constant term. 
This result coincides with the expression (\ref{26}) obtained through the twist field method,
once the relation $\epsilon = 2 \eta$ between the UV parameters is imposed. 
Thus, the method of \cite{Cardy-Tonni16} allows to recover the results of \cite{Dubail.17,Laguna.17},
once the possible spatial dependence in the UV cutoff is properly taken into account \cite{Dubail.17}, 
as mentioned in \S\ref{sec:EEinhom}.

The above analysis gives access also to the entanglement spectrum. 
Combining the expressions (\ref{lambda_j log}) and (\ref{lambda_j expanded}) with \eqref{33} and (\ref{31}),
for the eigenvalues of the reduced density matrix in the inhomogeneous models we find
\bea
\label{lambda_j log tilded}
- \log \lambda_{j}
&=&
- \left( \Delta_j - \frac{c}{24}\, \right)  \log ( e^{-2\pi^2/\widetilde{W}_{A} } )
+ \log \widetilde{\mathcal{Z}}_{\textrm{\tiny annulus}}
\\
\label{lambda_j expanded tilded}
&=&
\frac{c}{12}\, \widetilde{W}_{A}  +\log \big( \langle a | 0 \rangle  \langle 0 | b \rangle \big)
+ \left( \Delta_j - \frac{c}{24}\right) \frac{2\pi^2}{ \widetilde{W}_{A}} 
+O(\epsilon^r )\,,
\eea
where (\ref{31}) has been employed in the last step also to evaluate the neglected terms. 
By comparing (\ref{34}) and (\ref{lambda_j expanded tilded}), it is straightforward to realise that $- \log \lambda_{j} =  S_A/2$ for any $j$ at leading order
also in the inhomogeneous case.

The maximum eigenvalue corresponds to $\Delta_j =0$; therefore we have
\be
\label{lambda_max curved}
- \log \lambda_{\textrm{\tiny max}}
\,=\,
\frac{c}{12}\, \widetilde{W}_{A}  +\log \big( \langle a | 0 \rangle  \langle 0 | b \rangle \big)
-  \frac{\pi^2 c}{ 12\, \widetilde{W}_{A}} 
+O(\epsilon^r)\,.
\ee

The gaps in the entanglement spectrum are also interesting quantities to consider and 
they can be easily obtained from (\ref{lambda_j log tilded}), 
finding $\log \lambda_{j} - \log \lambda_{k} =
2\pi^2(\Delta_k - \Delta_j) / \widetilde{W}_{A} $.
As for the gaps with respect to the maximum eigenvalue, they read
\beq
\label{eek tilded} 
\mathcal{E}_k 
=
\log \lambda_{\textrm{\tiny max}} - \log \lambda_{k}
=
\frac{2\pi^2   \Delta_k }{\widetilde{W}_{A} }\,.
\eeq

From \eqref{34} and \eqref{eek tilded}, also for these inhomogeneous cases we obtain (see (\ref{consist rel1 flat}))
\beq
\label{consist rel1 tilded}
S^{(n)}_A   \, \mathcal{E}_k  
\,=\,
\frac{ \pi^2}{6} \left( 1 + \frac{1}{n}\, \right) \Delta_k + \dots\,,
\eeq
where the dots correspond to terms which vanish as $\epsilon \to 0$.
Let us remark that the r.h.s. of (\ref{consist rel1 tilded}) is independent of the inhomogeneity parameters.


\section{An entanglement contour for inhomogeneous systems} 
\label{sec:contour}

The contour for the entanglement entropies in lattice models has been introduced in \S\ref{sec:intro}.
Explicit constructions has been proposed in \cite{chen-vidal} for free fermions
and in \cite{cdt-17} for harmonic lattices (see also \cite{br-04, frerot-roschilde}).

In the continuum limit, the  conditions  \eqref{contour def lattice},  
that are necessary but not sufficient to define a proper contour function for the entanglement entropies, 
become respectively
\beq
S_A^{(n)} = \int_{A_\epsilon}  s_A^{(n)}(x) \, dx\,,
\;\;\qquad\;\;
s_A^{(n)}(x)\geqslant 0\,.
\label{co1} 
\eeq
Although these conditions (or their counterparts (\ref{contour def lattice}) in the lattice) 
does not allow to find the function $s_A^{(n)}(x)$ in a unique way, 
in \cite{cdt-17} it has been observed
that the analysis of \cite{Cardy-Tonni16} provides a natural candidate for the
contour function for the entanglement entropies in CFT,
restricted to the class of configurations considered in  \cite{Cardy-Tonni16}, 
which includes the one of our interest. 
In the following, by employing the results of \S\ref{sec:EHI}, 
we adapt the observation made in \cite{cdt-17} for the homogeneous models 
to the inhomogeneous systems discussed above.

Considering the configuration we are interested in, namely
the subsystem $A=(x_0, L)$ in the segment $(-L,L)$,
the integration domain in (\ref{co1}) is  the interval $A_\epsilon = (x_0 + \epsilon, L)$.

In order to construct the contour function for the entanglement entropies 
of this configuration in inhomogeneous models, 
from \eqref{10}, we notice that the expression \eqref{33} can be written as 
\beq
\widetilde{W}_A  = \int_{\tilde{A}_{\tilde{\epsilon}}}   \tilde{f}'( \tilde{x}) \, d \tilde{x}
= 
\tilde{f}(\tilde{L})- \tilde{f}(\tilde{x}_0 + \tilde{\epsilon}) \,.
\label{32}
\eeq
By employing this observation in (\ref{34}), we find for the entanglement entropies that
\be
S_A^{(n)} 
= \frac{c}{12} \left( 1 + \frac{1}{n} \right)   \int_{A_\epsilon}  \tilde{f}'(\tilde{x}(x)) \, \tilde{x}'(x)\, dx\,,
\label{co2} 
\ee
up to a additive constant term. 
Then, by comparing \eqref{co1} and \eqref{co2}, we are naturally lead to the following candidate for the contour function of the entanglement entropies
\beq
s_A^{(n)}(x) = \frac{c}{12} \left( 1 + \frac{1}{n} \right)  {\cal S}_A(x) + \frac{C_n}{\ell }\,,
\label{co3} 
\eeq
where the function $ {\cal S}_A(x)$ is defined as follows
\beq
 {\cal S}_A(x) =  \tilde{x}'(x) \,   \tilde{f}'(\tilde{x}(x))\,,
 \label{co3bis} 
\eeq
and $C_n$ is the non universal constant term in \eqref{5}.
We find it worth remarking that the contour function (\ref{co3}) does not depend on the 
UV cutoff, which plays a crucial role in the determination of the
spatial dependence of the entanglement entropies in the specific model, as mentioned in \S\ref{sec:EEinhom}.

Let us consider the behaviour of the contour function (\ref{co3}) as $x \to x_0^+$.
This can be easily done by exploiting the similar analysis made for the weight function (\ref{30b}), whose result is (\ref{beta expand x0}),
which can be written also in a form involving the Weyl factor reported in the text below (\ref{beta expand x0}).
From the latter observation we find that 
\beq
\label{contour linear x0}
\mathcal{S}_A(x) = \frac{1}{x - x_0} 
+ e^{-\sigma(x_0)} \tilde{f}_0 - \frac{\sigma'(x_0)}{2} 
+ O\big((x-x_0)\big)\,,
\eeq
where $\tilde{f}_0$ has been introduced in the text below (\ref{beta expand x0}).
The leading term in (\ref{contour linear x0}) is responsible of the logarithmic divergence in the entanglement entropies as $\epsilon \to 0$.
Thus, the contour function that we have constructed can quantify the expectation that 
the entanglement between two regions is mainly due to the parts of the complementary regions close to the entangling point
separating $A$ and $B$.

It is interesting to mention that, by comparing \eqref{30b} and \eqref{co3bis}, one finds the following relation 
\beq
\beta_A(x) \, {\cal S}_A(x) = 1\,,
\label{co4}
\eeq 
which can be expressed also  in terms of the 
contour function for the entanglement entropies (\ref{co3}).
Indeed, by introducing the proper factors from (\ref{1}), (\ref{30a}) and (\ref{co3}), 
we have that (\ref{co4}) leads to
\beq
 \big[ 2\pi \beta_A(x)\big]
s^{(n)}_A(x)
= 
\frac{ \pi}{6} \left( 1 + \frac{1}{n} \,\right) c \, + \dots\,,
\label{co5}
\eeq 
where the dots corresponds to infinitesimal terms as $\epsilon \to 0$.
The relation (\ref{co5}) exhibits an intriguing similarity with (\ref{consist rel1 tilded}), with the crucial difference that
the quantities involved in the l.h.s. of (\ref{co5}) depend on the position $x \in A$.


\section{Numerical results  for the rainbow chain} 
\label{sec:numerical}

The analytic expressions discussed in \S\ref{sec:EHI} and \S\ref{sec:contour} hold 
for a generic CFT in the static background given by (\ref{14}). 

In this section we focus on the rainbow model. 
Considering rainbow chains defined by (\ref{eq:rainbow_ham}),
we present numerical confirmation 
of the analytic expressions of \S\ref{sec:EHI} and \S\ref{sec:contour} for this model. 
In particular, for the configuration given by a block $A$ adjacent to the boundary of a segment,
we compute the entanglement hamiltonian, the entanglement spectrum and a contour for the entanglement entropies.

\subsection{Entanglement hamiltonian}
\label{sec-eh-rb}

In order to estimate the entanglement hamiltonian for a block of our system,
we have developed an {\em ab initio} numerical procedure similar to 
the one based on {\em machine learning} techniques described in \cite{Fujita.17}.
A different approach 
has been proposed in \cite{Cirac.11}.

The starting point of our numerical procedure is the following ansatz for the entanglement hamiltonian
\beq
H_A(\bbeta) 
\,=\, -\, \frac{1}{2} \sum_{i\, \in \,A}
\beta_i\, d^\dagger_i d_{i+1} + \textrm{h.c.}\,,
\label{eq:entham}
\eeq
where $\bbeta=\{\beta_i\, , i\in A\}$ is the vector whose elements are the couplings of the entanglement hamiltonian.
We remark that the ansatz (\ref{eq:entham}) does not capture the complete entanglement hamiltonian of the underlying lattice model.
Indeed, considering the detailed results obtained in \cite{ent-ham-latt, peschel-eisler-17} 
for  a single block in a homogenous chain of free fermions on the infinite line,
in the entanglement hamiltonian non vanishing couplings occur also between sites that are not nearest neighbours.
Nonetheless, these couplings are expected to be very small, therefore we are allowed to neglect them as first approximation.

In order to impose that the elements of the vector $\bbeta$ 
provide a density matrix $\rho_A(\bbeta) \equiv e^{ - H_A(\bbeta)}$ 
approximating the reduced density matrix $\rho_A$ for the subsystem $A$,
we exploit the Wick's theorem, 
which ensures that $\rho_A$ is characterised by
the correlation matrix $C_{ij}\equiv \<c^\dagger_i c_j\>$ 
restricted to the block $A$, i.e. with $i$, $j\in A$.
Thus, given a choice of  $\bbeta$, let us introduce 
\beq
C_{ij}(\bbeta) \equiv \Tr\big(\rho_A(\bbeta)\, d^\dagger_i d_j\big)\,,
\;\;\qquad\;\;
i,j \,\in\, A\,,
\label{eq:reduced_dm}
\eeq
with $\rho_A(\bbeta)$ defined through (\ref{eq:entham}),
and also the following error function
\beq
E(\bbeta)\equiv \sum_{i,j\,\in\, A} \big[C_{ij}-C_{ij}(\bbeta)\big]^2\,.
\label{eq:error_beta}
\eeq
The numerical values for the elements of $\bbeta$ are obtained by minimising 
this error function through standard optimization techniques (Powell method) \cite{NRC}. 
The optimization procedure begins with a homogeneous seed and then seeks a value of $\bbeta$
which fits the exact correlation matrix within a certain tolerance, which is fixed to $10^{-6}$ in the worst case. 
Our algorithm was able to find a solution for all the cases that we have explored.

In the continuum limit, by comparing (\ref{1}) with (\ref{eq:entham}) , 
we expect that $H_A(\bbeta) \to 2\pi K_A$,
where $K_A$ is the entanglement hamiltonian \eqref{30a}.
For a Dirac fermion, the component $T_{00}(x)$ of the energy-momentum tensor is given by
\beq
T_{00}(x) \,=\,
 \iu \left( \psi^\dagger_R(x) \partial_x \psi_R(x) - \psi^\dagger_L(x) \partial_x \psi_L(x) \right) .
\label{t00}
\eeq
Thus, for the couplings $\beta_i$ in (\ref{eq:entham}) we expect that $\beta_i \to 2\pi\beta_A(x)$ in the continuum limit.

The weight function $\beta_A(x)$ to adopt in \eqref{30a} for the
rainbow model can be constructed 
by specifying the general expression
\eqref{ent temp xtilde} for the map $\tilde{x}(x)$ characterising the
rainbow model, given in \eqref{18}.
This leads to the following expression 
\beq
\beta_A(x)
\;=\;
\frac{2(e^\lambda - 1)L}{\pi \lambda}
\;\,
\frac{\sin\big(\tfrac{\pi}{2} \,\tfrac{(e^{\lambda |x/L|} - 1) \,\textrm{sign}(x/L) }{e^\lambda -1} \big)
-
\sin\big(\tfrac{\pi}{2} \,\tfrac{(e^{\lambda |x_0/L|} - 1) \,\textrm{sign}(x_0/L) }{e^\lambda -1} \big)
  }{
  e^{\lambda |x/L|}\, 
\cos\big(\tfrac{\pi}{2} \,\tfrac{(e^{\lambda |x_0/L|} - 1) \,\textrm{sign}(x_0/L) }{e^\lambda -1} \big)
  }\,.
\label{ent temp exp}
\eeq
From this formula it is straightforward to observe that $\beta_A(x)/L$ is a function of $x/L$, 
where $\lambda=hL$ and $x_0/L$ enter as the parameters characterising the subsystem $A$.
In the homogeneous case, namely for $h=0$, we have that \eqref{ent temp exp} reduces to \eqref{weight function homo}, 
as expected.
Also the expansion (\ref{beta expand x0}) of $\beta_A(x)$ for $x \to x_0^+$ can be specified for the rainbow model,
but we do not report the resulting expression.
Let us stress that the features of the model occur in the subleading $O((x-x_0)^2)$ term of (\ref{beta expand x0}).

We find it more instructive first to discuss \eqref{ent temp exp} for some interesting special regimes.

\begin{figure}[t!]
\vspace{.2cm}
\begin{center} 
 \includegraphics[width=.7\textwidth]{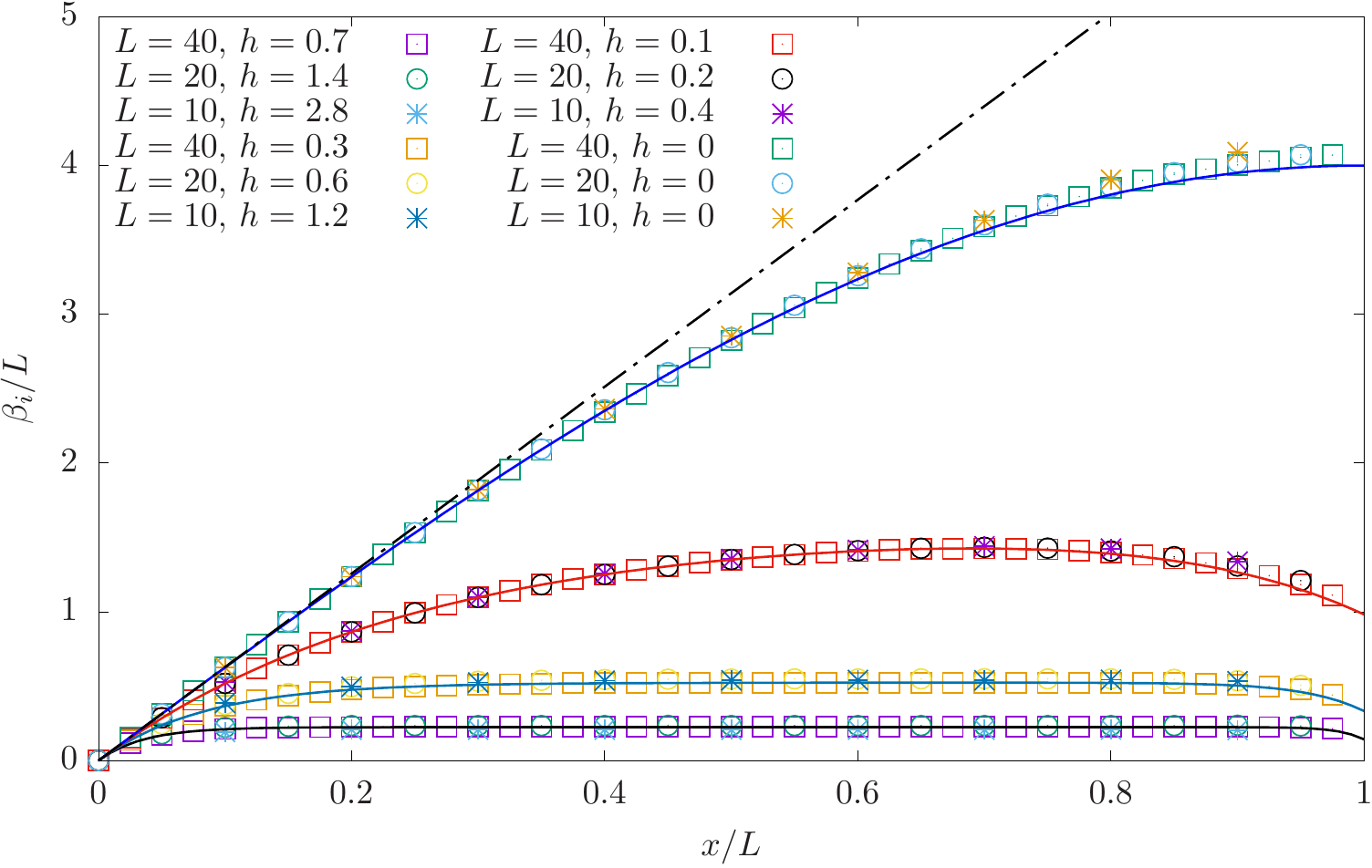}
 \caption{
   Couplings $\beta_i/L$ in the entanglement hamiltonian (\ref{eq:entham})
   corresponding to the right half $A=(0,L)$, i.e. for $x_0=0$, of various rainbow chains,
   obtained numerically as discussed in \S\ref{sec-eh-rb}.
   The data points agree with the analytic expression of $2\pi\beta_A(x)/L$ in (\ref{beta2}), 
   which is a function of $x/L$ parameterised by $\lambda=hL$.
   The dashed-dotted straight line has slope $2\pi$ and highlights the behaviour inherited from the 
   Bisognano-Wichmann theorem near the entangling point $x_0=0$.
   When $\lambda \gg 1$ a plateau occurs which provides the thermofield double interpretation 
   for the ground state of the rainbow chain in the strong coupling regime (see (\ref{eq:tfd})). 
   }
     \label{fig:beta1}
\end{center}   
\end{figure}

When the block $A$ is half of the rainbow chain,
the weight function in the entanglement hamiltonian can be obtained by setting $x_0 = 0$ in \eqref{ent temp exp}.
The result reads
\beq
\beta_A(x) 
\,\equiv\,
\frac{ 2 L}{\pi}   \;
\frac{e^{\lambda} -1}{ \lambda }
\;e^{-\lambda x/L}\;
\sin\bigg(  \frac{\pi}{2}  \; \frac{e^{\lambda x/L} -1}{e^{\lambda} -1} \bigg) \,,
\label{beta2} 
\eeq
with  $x \in (0,L)$, which is manifestly positive. 
Taking $\lambda \gg 1$ in  \eqref{beta2}, we find 
\beq
\beta_A(x)  \,=\, \frac{1}{h} \big( 1 - e^{- h x} \big) + \dots\,,
\label{beta2 large lambda} 
\eeq
where the dots denote subleading terms. 
The expansion \eqref{beta2 large lambda} captures the main features of this regime; 
indeed, $\beta_A(x) \simeq x$ for $x \ll 1/h$ and $\beta_A(x) \simeq 1/h$ for
$x \gg 1/h$; namely $\beta_A(x)$ is linear near to the entangling point and becomes flat for a large part
of the remaining interval.

 In Fig.\,\ref{fig:beta1} we show the numerical values of the coefficients $\beta_i/L$ 
for the right half (i.e. $x_0 = 0$) of various rainbow chains  as a function of $x/L$ (with $i=x$), 
along with the theoretical prediction $2\pi\beta_A(x)/L$ in the continuum limit given by \eqref{beta2},
which is parameterised by the combination $\lambda=hL$.
An excellent agreement is observed between the numerical data and the CFT formula
along the entire subsystem $A$.
For $x \simeq x_0^+$, notice the peculiar behaviour of the Bisognano-Wichmann 
theorem, which corresponds to the dashed and dotted line in Fig.\,\ref{fig:beta1}, 
as expected from (\ref{beta expand x0}).

In the homogeneous case, i.e. for $h=0$, the data for $\beta_i/L$ in Fig.\,\ref{fig:beta1} follow the CFT curve \eqref{weight function homo}.
As $\lambda$ increases, the $\beta_i/L$ develops a plateau at a height $2\pi / \lambda$,
in agreement with the prediction \eqref{beta2 large lambda}.

\begin{figure}[t!]
\vspace{.2cm}
  \center
    \includegraphics[width=.7\textwidth]{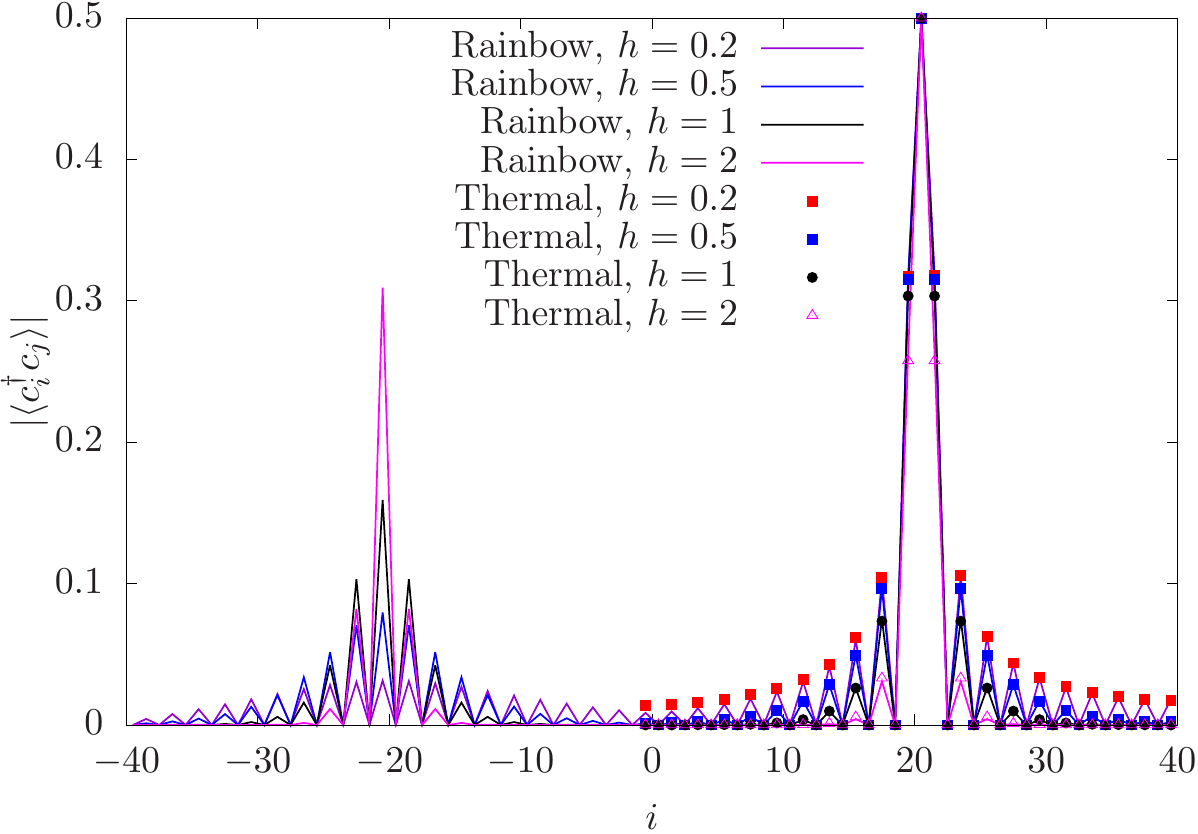}
  \center
  \caption{
  Correlation matrix for the rainbow state compared to the
    thermofield double approximation. For different rainbow systems
    with $L=40$, we show $|\langle c^\dagger_i c_j \rangle|$, with
    $j=20$. 
    The points having $i=1, \dots, 40$ have been computed using a thermal density matrix of a hamiltonian
    with uniform couplings and temperature (\ref{tr}). 
    The continuos lines correspond to the exact correlations in the rainbow ground state,
    which have been evaluated for the entire chain. 
    Precise agreement is found within the block $A$ between the two different computations of the correlator.  
    A peak occurs also at the site $i=-20$, which is entangled with the site $i=20$
    because of a long distance bond.}
  \label{fig:tfd}
\end{figure}

An interesting consequence of \eqref{beta2 large lambda} is that we can approximate the reduced density matrix for $\lambda\gg 1$ as follows
\beq
\rho_A \simeq {\rm exp} \left(  - \, \frac{2 \pi}{h} \int_A  T_{00}(x) \,dx\right)  ,
\label{eq:tfd}
\eeq
which is the thermal density matrix of a CFT with $c=1$ and an effective temperature given by \cite{Ramirez.15}
\beq
T_R = \frac{h}{2 \pi}\,.
\label{tr}
\eeq
This result is consistent with the entanglement entropy \eqref{28} in this regime, 
which becomes $S_A \simeq h L/6$ for $n=1$, 
namely the thermal entropy of a CFT with $c=1$ at finite temperature \eqref{tr}.

The latter observation about the entanglement entropy led to the
interpretation of the ground state of the rainbow model as a
thermofield double \cite{Ramirez.15}, that is a pure state
defined on the tensor product $\mathcal{H}_l \otimes \mathcal{H}_r$ 
of two copies of the same Hilbert space associated to a CFT 
and whose Schmidt decomposition is given by
\beq
\ket|\psi> = 
\sum_n e^{ -  \frac{1}{2} \beta_R E_n} \, \ket|n>_l\,\ket|n>_r\,,
\;\;\qquad \;\;
\beta_R = \frac{1}{T_R} \,,
\label{eq:tfd_state}
\eeq
where $\ket|n>_l$ and $\ket|n>_r$ are the eigenstates of a CFT
hamiltonian corresponding to the energy level $E_n$ for the left and
right blocks of the system. Tracing out either the left or the right
half of the system gives $\sum_n e^{-\beta_R E_n} \ket|n>\bra<n|$,
which is the thermal density matrix  of a single CFT with inverse temperature $\beta_R$.

A consequence of the thermofield double structure of the ground state of the rainbow
chain is that the expectation values of operators belonging to the
same half chain can be obtained as averages on half chain 
at finite temperature $T_R$ given by  \eqref{tr}. 
Considering, for instance, the correlator 
$ \langle c^\dagger_i c_j \rangle 
$, in Fig.\,\ref{fig:tfd} we show the
values of $|\langle c^\dagger_i c_j \rangle|$ for a chain with $2L=80$
sites, $j=20$ and $i=1, 2, \dots, 40$.  
Taking the right half chain as subsystem, for $i\geqslant 1$ 
this correlator can be computed also as a correlator of a uniform chain 
made by $40$ sites at finite temperature $T_R$.
An excellent agreement is observed between these two ways to compute this correlator.

\begin{figure}[t!]
\vspace{.2cm}
  \begin{center}
  \includegraphics[width=.7\textwidth]{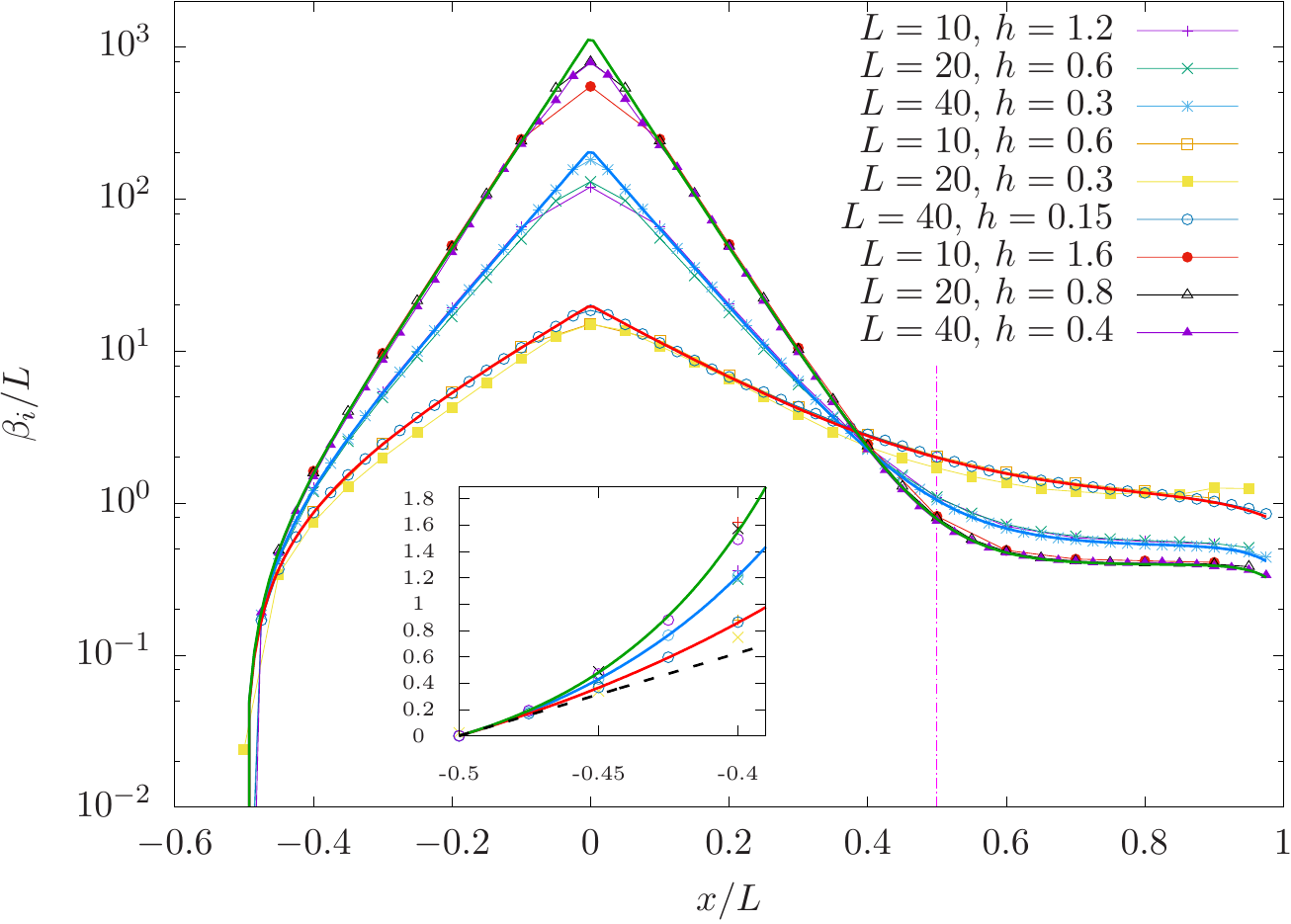}
  \caption{
     Couplings $\beta_i/L$ in the entanglement hamiltonian (\ref{eq:entham})
   for the block $A=(x_0,L)$ with $x_0 =-L/2$ in various rainbow chains,
   obtained numerically as discussed in \S\ref{sec-eh-rb}.
   The continuos solid lines correspond to the analytic prediction 
   for $2\pi\beta_A(x)/L$ given by (\ref{ent temp exp}) in terms of $x/L$.
   The values of $L$ and $h$ have been chosen in order to 
   highlight the collapse of the numerical data having the same $\lambda=hL$.
   The decay from $x=0$ outwards is approximately exponential.
   The vertical dashed line, marks the boundary between the active and inactive regions. 
    Inset: zoom for $x\sim x_0^+$, which emphasizes the Bisognano-Wichmann behavior 
    (dashed straight line with slope $2\pi$) near the entangling point. 
    }
  \label{fig:beta3}
    \end{center} 
\end{figure}

Finally, let us consider blocks $A=(x_0,L)$ having $x_0 \neq 0$. The
general expression for the weight function $\beta_A(x)$ in the
entanglement hamiltonian \eqref{30a} is given by \eqref{ent temp exp},
which is a positive function for $x\in A$, as discussed below (\ref{30b}).
Taking the limit $\lambda \gg 1$ of  \eqref{ent temp exp} with fixed $x/L$, we find 
\beq
\label{ent tem rb large}
\beta_A(x)
\, \simeq \,
\frac{\textrm{sign}(x) - \textrm{sign}(x_0) \,e^{-h(|x|- |x_0|)} }{h}\,,
\eeq
where we have reported only the leading terms. 
When $x_0 > 0$ we get that $\beta_A(x) \simeq (1 - e^{-h(x- x_0)})/h$, which reduces to
\eqref{beta2 large lambda} for $x_0 = 0$, as expected. The
qualitative behaviour of $\beta_A(x)$ for $x_0 > 0$ is very similar to the one
observed for the $x_0 = 0$ case. 
Instead, when $x_0 < 0$ the expression \eqref{ent tem rb large} becomes
\beq 
\label{beta4}
\beta_A(x) 
\,\simeq\,
\left\{\begin{array}{ll} 
e^{h(|x_0|-|x|)}/h \hspace{1cm} &  - |x_0| < x < |x_0|\,,
\\ 
\rule{0pt}{.6cm} 
1/h & |x_0| < x < L \,,
\end{array}\right.
\eeq
which displays qualitative different features; 
indeed an exponential behavior is observed in the region $|x|<|x_0|$ 
and the usual plateau $\beta_A(x)\simeq 1/h$ occurs for $x>|x_0|$. 

An intuitive explanation of this behaviour is the following.
The rainbow state in the strong coupling limit is composed of
long-distance valence bonds symmetrically placed around the origin,
between sites at $+x$ and $-x$ (see Fig.\,\ref{fig:RB_scheme}). 
Thus, in this limit it is natural to identify two regions of the interval $A=(x_0,L)$, with $x_0<0$: 
an inactive zone $(-|x_0|,+|x_0|)$ containing bonds which do not leave the block, 
and an active zone $(|x_0|,L)$, which contains bonds linking the block $A$ to its complement $B$. 
The vertical dashed line in Fig.\,\ref{fig:beta3} separates these two regions in $A$. 
For any finite $h$, fluctuations in the active zone are higher than the ones in the inactive zone 
because of the broken bonds. 
Thus, in the limit of large $\lambda$, the coefficients of the entanglement hamiltonian 
present the usual plateau of height $1/h$ in the active zone, while in the
inactive zone they present an exponential decay from the origin which is similar to
the one occurring in the original Hamiltonian. 

The fact that the couplings $\beta_i$ take very high values in the inactive zone 
(the peak is reached at the origin, where $\beta_A(0) \propto \exp(h|x_0|)$) 
inhibits fluctuations in that zone. Indeed, in Fig.\,\ref{fig:beta3} we can observe that our numerical
technique to estimate $\beta_i$ makes larger errors near $x=0$. 
This happens because the high values of the couplings $\beta_i$ in the inactive region cause the
elements of the correlation matrix to be rather insensitive to details on the $\beta_i$'s.

\subsection{Entanglement spectrum}

The reduced density matrix $\rho_A$ of a free fermion system
decomposes into the product of single body density matrices
\cite{ent-ham-latt}
\beq
\rho_A \,= \,  e^{-\sum_p \varepsilon_p \, \hat{n}_p - r_0}  \,,
\label{37}
\eeq 
where $\hat{n}_p$ are the occupation number operators, whose eigenvalues are either 0 or 1, 
and the corresponding $\varepsilon_p$ are called single body entanglement energies, 
which are related to the eigenvalues  $\nu_p = \< \hat{n}_p \>$ of the block correlation matrix 
$C_{ij}=\<c^\dagger_i c_j\>$ with $i,j\in A$ as follows
\beq
\nu_p \,=\, 
\frac{1}{\exp(\varepsilon_p) + 1}\,.
\label{38}
\eeq
The normalization condition $\Tr \rho_A = 1$ for (\ref{37}) provides the constant $r_0$.
The eigenvalues of $\rho_A$ in (\ref{37}) can be written as $\lambda_j \equiv e^{-E_j} \in (0,1)$,
in terms of the entanglement energies $E_j$,
which can be obtained as $E_j = \sum_{p} \varepsilon_p n_p + r_0$, 
where the index $j$ denotes the set $\{ n_p \}$ of the occupation numbers providing $E_j$.

\begin{figure}
\vspace{.3cm}
  \hspace{-.9cm}
  \includegraphics[width=.52\textwidth]{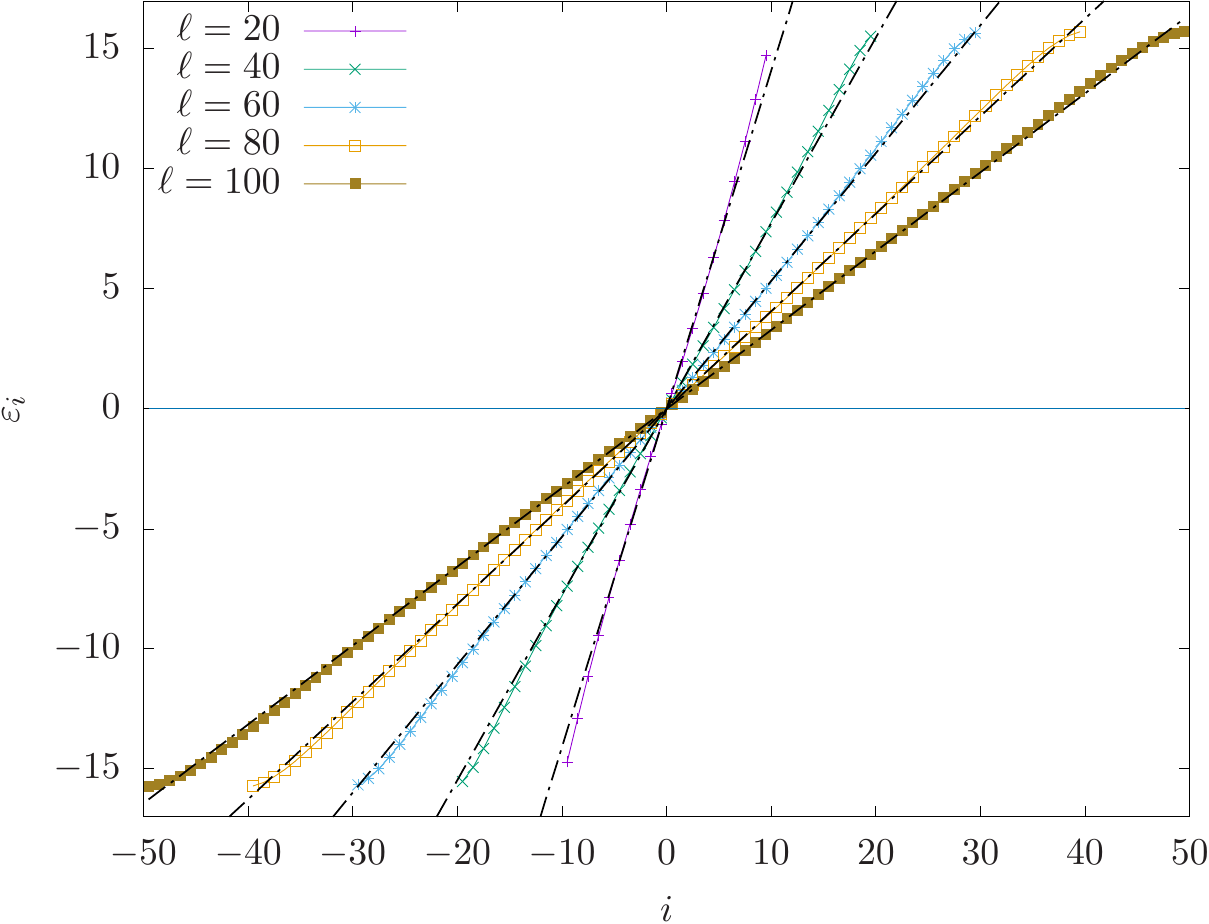}
  \hspace{.2cm}
  \includegraphics[width=.52\textwidth]{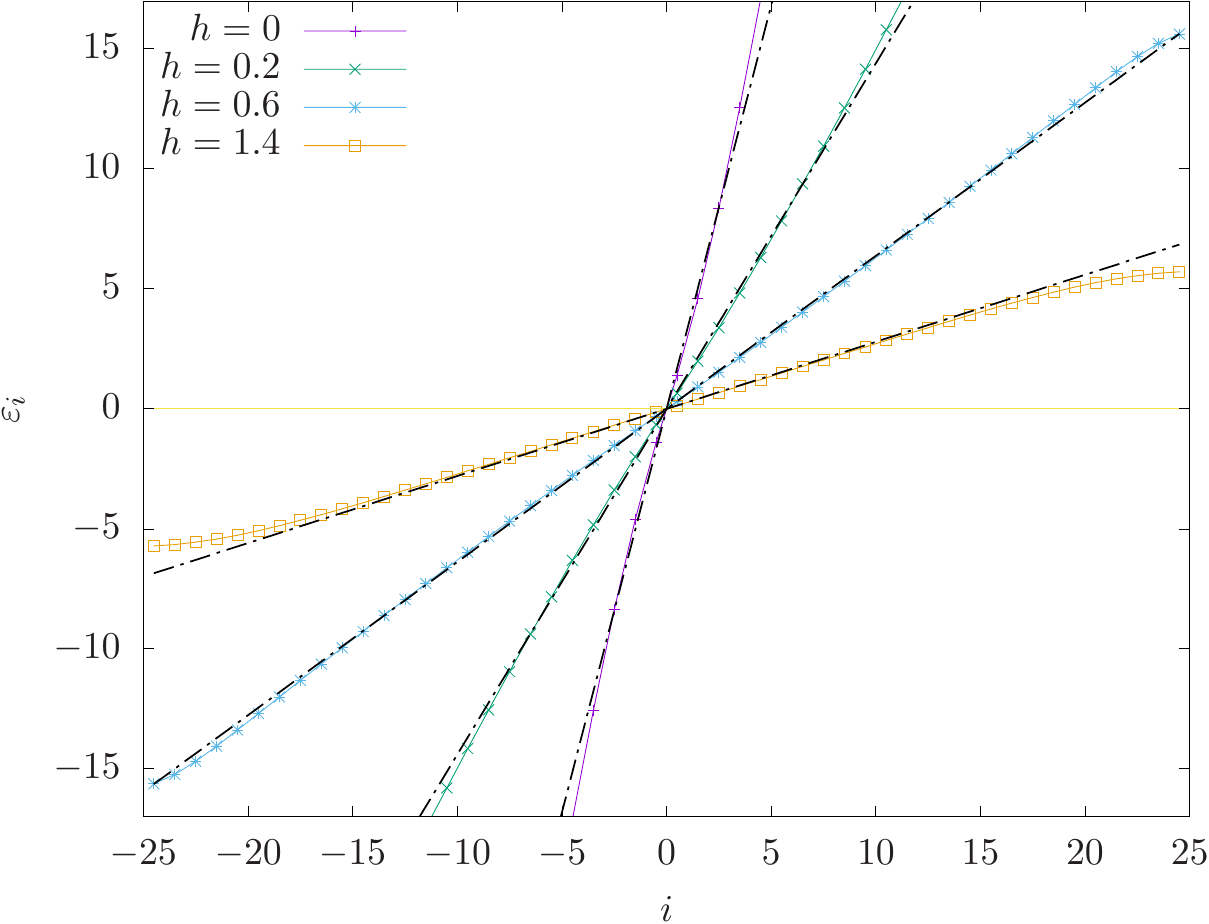}
  \caption{ 
    Single body entanglement energies for a rainbow systems with $L=100$.
    Left: 
    Rainbow chain with $h=1.4$ fixed and different values of the block size $\ell$, from 20 up to 100.
    Right: 
    Rainbow chain with different values of $h$ and the block $A$ made by half chain.
   An approximately linear behavior is observed in all cases,
   confirming the theoretical prediction given by (\ref{40}) and \eqref{41}.
   }
  \label{fig:es}
\end{figure}

In Fig.\,\ref{fig:es} we show the single body entanglement energies
$\varepsilon_p$ for a rainbow chain made by $2 L=200$ sites and
blocks adjacent to a boundary whose length varies between 
$\ell =10$ sites and $\ell=100$ sites. 
For even sizes $\ell$, the different $\varepsilon_p$
are labeled by $p = \pm \frac{1}{2}, \pm \frac{3}{2}, \dots, \pm
\frac{\ell -1}{2}$. 
The spectrum exhibits a particle-hole symmetry
$\varepsilon_{-p} = -\varepsilon_p$ and it is approximately linear
for small values of $p$.
In particular, $\varepsilon_p \propto p + o(p^3)$ \cite{Ramirez.15}. 
This behaviour corresponds to a massless free fermion with open boundary conditions.

The scaling dimensions $\Delta_j$ occurring in the CFT formulas
\eqref{lambda cft flat} and (\ref{lambda_j log tilded}) for these free fermion models
can be identified with the sums of the half-odd integers $|p|$.  
For instance, the lowest entanglement energy 
$E_{0} = \sum_{p=-\frac{\ell-1}{2}}^{-\frac{1}{2}} \varepsilon_p + r_0$ 
is obtained by filling all the negative single body energy levels. 
The next entanglement energy in the same sector is given by 
$E_1 = E_0 + \varepsilon_{\frac{1}{2}} - \varepsilon_{-\frac{1}{2}}$, 
being the corresponding state obtained through a particle-hole excitation.

\subsubsection{Largest eigenvalue of the reduced density matrix.}

An interesting quantity to consider is the largest eigenvalue $\lambda_{\textrm{\tiny max}} = e^{- E_0} $ of the density matrix,
which is related to the lowest entanglement energy $E_0$ and provides the single-copy entanglement \cite{single-copy-ent}.

For a free fermion chain, by employing the above discussion, 
we find that $\lambda_{\rm max}$ can be computed as follows
\beq
\lambda_{\textrm{\tiny max}} 
\,=\,
  \frac{ \prod_{p < 0} e^{- \varepsilon_p} }{ \prod_p ( 1 + e^{-\varepsilon_p} ) } 
\,=\,
 \prod_{p < 0} \nu_p \, \prod_{p >0} ( 1 - \nu_p)  \,.
\label{lmax}
\eeq

\begin{figure}[t!]
\vspace{.2cm}
\hspace{-.8cm}
\includegraphics[width=7.9cm]{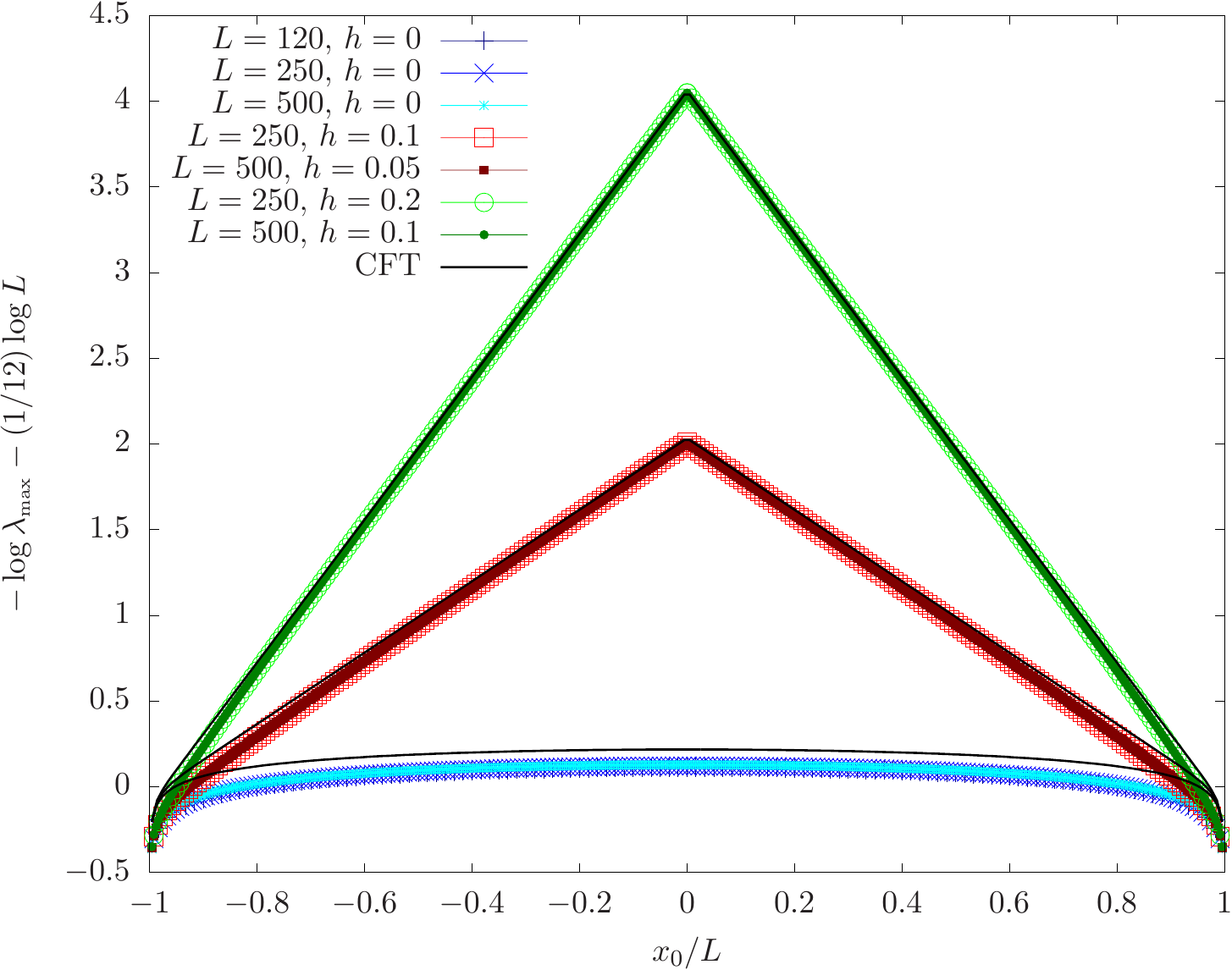}
\hspace{.1cm}
\includegraphics[width=9cm]{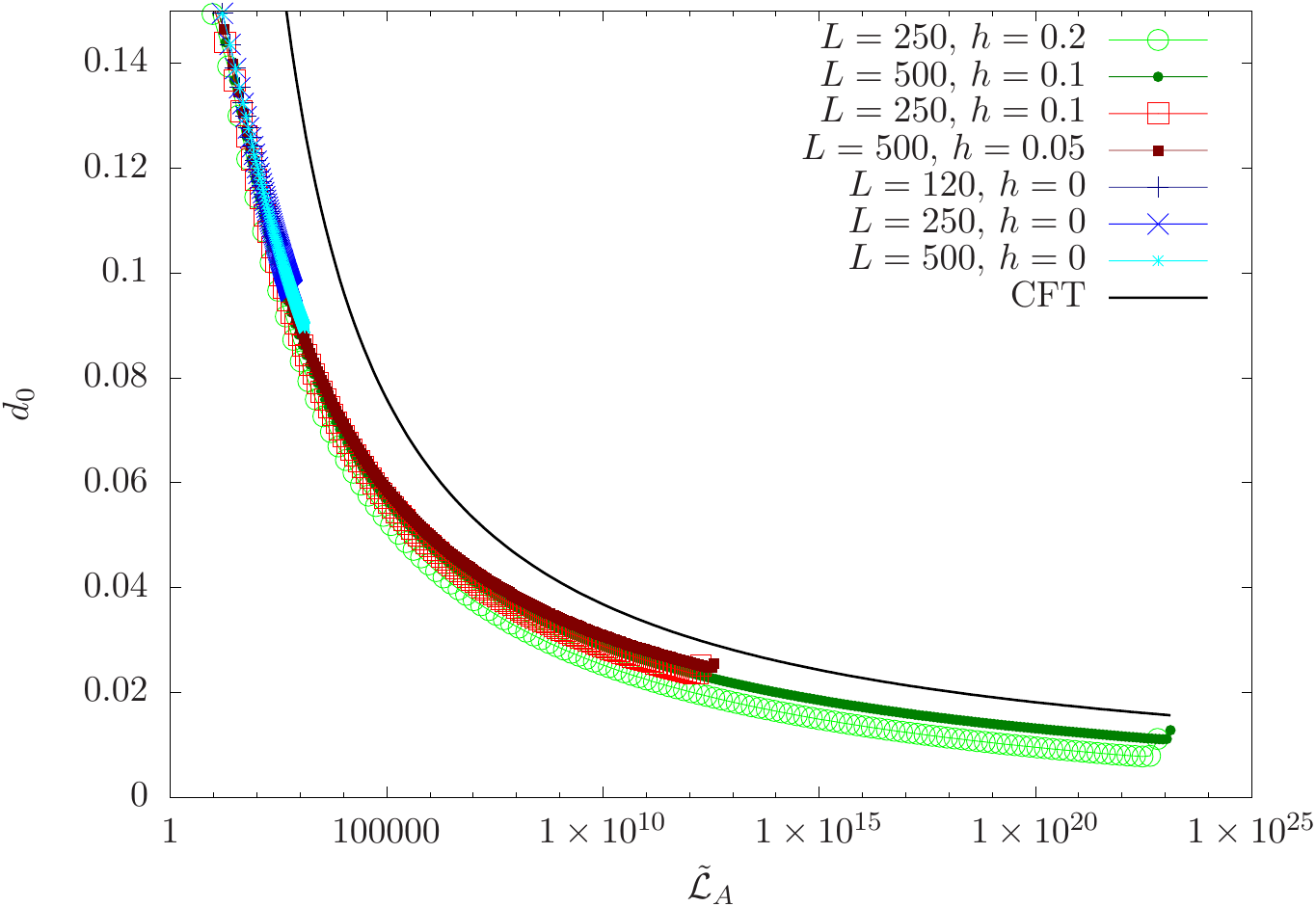}
  \caption{
  Left: Largest eigenvalue $ \lambda_{\textrm{\tiny max}}$ of the reduced density matrix for rainbow systems 
  whose size $L$ and parameter $h$ have been chosen in order to highlight the collapse of the numerical data having the same $\lambda$.
  The solid lines correspond to the theoretical prediction \eqref{lambda_max sub}. 
  The slight mismatch for $\lambda  =0 $ is expected to decrease for higher values of $L$ (as shown also in the right panel).
  Right: The combination (\ref{d0_def}) in terms of the effective length (\ref{tLA})
  The solid line corresponds to the CFT estimate (\ref{d0cft_def}) for the subleading correction, 
  which agrees with the numerical data up to a constant shift. 
  }
  \label{fig:largest}
\end{figure}

For a general system, the largest eigenvalue $\lambda_{\textrm{\tiny max}}$ can be obtained by taking the limit $n \to +\infty$ 
of the R\'enyi entropies, which gives $S_A^{(n)} \to  -\log \lambda_{\textrm{\tiny max}}$\,.
In the rainbow CFT model and for an interval $A=(x_0, L)$ adjacent to the boundary of the segment
$(-L, L)$, we can take the limit $n \to +\infty$ of the result found in \cite{Laguna.17}, which gives
\be
\label{lambda_max large}
-\log \lambda_{\textrm{\tiny max}}
\,=\,
\frac{1}{12}\,
\log \tilde{\mathcal{L}}_A 
+
\frac{Q_\infty}{2} \,,
\ee
where we have introduced the following effective length
\beq
\tilde{\mathcal{L}}_A 
\equiv 
\frac{8 \tilde{L}}{\pi}\,
e^{\sigma(x_0)}\,   \,\sin \bigg( \frac{\pi \tilde{\ell}}{2\tilde{L}}\bigg) 
\,=\,
 \frac{8}{\pi}\,
e^{-h|x_0|}\; \frac{e^{hL} - 1}{h}\; \cos \left( \frac{\pi}{2}\;\frac{e^{h|x_0|} - 1} {e^{hL} - 1}  \right)  
\,\equiv\,
L \, \tilde{L}_A\,,
\label{tLA} 
\eeq
being $\tilde{L}_A$ the function of $\lambda$ and $x_0/L$ given by
\beq
\tilde{L}_A \equiv 
 \frac{8}{\pi}\,
 e^{-\lambda |x_0/L|}\; \frac{e^{\lambda} - 1}{\lambda}\; \cos \left( \frac{\pi}{2}\;\frac{e^{\lambda |x_0/L|} - 1} {e^{\lambda} - 1}  \right).
 \label{eq:tildeLA}
\eeq
The numerical value of the constant $Q_\infty$ in (\ref{lambda_max large}) reads $Q_\infty = 0.2797$,
which has been obtained by taking the limit $n \rightarrow +\infty$ of the constants $Q_n$ 
introduced in \cite{JK04} for the XX model (see also \cite{calabrese-essler,fagotti-calabrese}).

By using (\ref{lambda_max large}) and (\ref{tLA}), it is straightforward to construct the following combination
\beq
\label{lambda_max sub}
 -\log \lambda_{\textrm{\tiny max}} - \frac{1}{12} \log L
\,=\,
\frac{1}{12}\, \log  \tilde{L}_A
+ 
\frac{Q_\infty}{2} \,.
\eeq
This analytic expression has been compared against the numerical values obtained from the lattice through (\ref{lmax})
in the left panel of Fig.\,\ref{fig:largest}.
The agreement is very good and it improves as $\lambda$ increases. 
In the homogenous case, i.e. for $\lambda =0$, a slight deviation is observed between the data points and the corresponding 
analytic curve. 
We expect that higher values of $L$ are needed in order to improve the matching with the CFT curve. 

The subleading corrections to (\ref{lambda_max large}) can be analysed by considering
\be
\label{d0_def}
d_0 \equiv 
 -\log \lambda_{\textrm{\tiny max}} 
 - \left( 
 \frac{1}{12}\, \log  \tilde{\mathcal{L}}_A 
+ 
\frac{Q_\infty}{2}
 \right),
\ee
whose data point are shown in the right panel of Fig.\,\ref{fig:largest}.
From the data corresponding to $\lambda=0$, it is evident that
the agreement with the CFT curve improves as $L$ increases. 
Although the subleading corrections (\ref{d0_def}) strongly depend on the underlying model
(see e.g. \cite{calabrese-essler,fagotti-calabrese} for homogenous cases),
in the right panel of Fig.\,\ref{fig:largest} we have compared the data point with the 
following analytic curve 
\be
\label{d0cft_def}
d_0^{\textrm{\tiny CFT}} 
=
-\,\frac{\pi^2}{12 \,\log (\tilde{\mathcal{L}}_A /2)}\,,
\ee
which has been obtained from (\ref{lambda_max curved}) with $c=1$ and (\ref{tLA}).
The qualitative behaviour of the lattice data in the right panel of Fig.\,\ref{fig:largest}
is nicely captured by (\ref{d0cft_def}), except for a vertical shift, 
which could be attributed to the fact that the $O(1)$ term in (\ref{lambda_max curved})  is not universal. 
%

\subsubsection{First gap in the entanglement spectrum.}
\label{sec:gap1rb}

Interesting quantities which characterise the entanglement spectrum for a particular configuration are
the gaps between the eigenvalues of the reduced density matrix. 
In the CFT analysis of \S\ref{sec:EHI} the gaps with respect to the largest eigenvalue 
are provided by (\ref{eek tilded}) for the inhomogeneous models that we are considering.

Given the ordered sequence $E_0 < E_1 < E_2 < \dots$ of the entanglement energies, 
in the following we focus on the first gap in the entanglement spectrum, namely
\beq
{\cal E}_1 \equiv E_1 - E_0  =2  \varepsilon_{\frac{1}{2}}\,,
\label{40}
\eeq
where in the last step the relation 
$\varepsilon_{-\frac{1}{2}} = -\varepsilon_{-\frac{1}{2}}$ has been used. 
The gap ${\cal E}_1$ gives also the slopes of the straight dashed-dotted lines in Fig.\,\ref{fig:es}. 

The linearity of the single body entanglement energies $\varepsilon_i$
implies that the other gaps ${\cal E}_k$ are multiples of ${\cal E}_1$.

\begin{figure}[t!]
\vspace{.2cm}
\hspace{-1.2cm}
  \includegraphics[width=.585\textwidth]{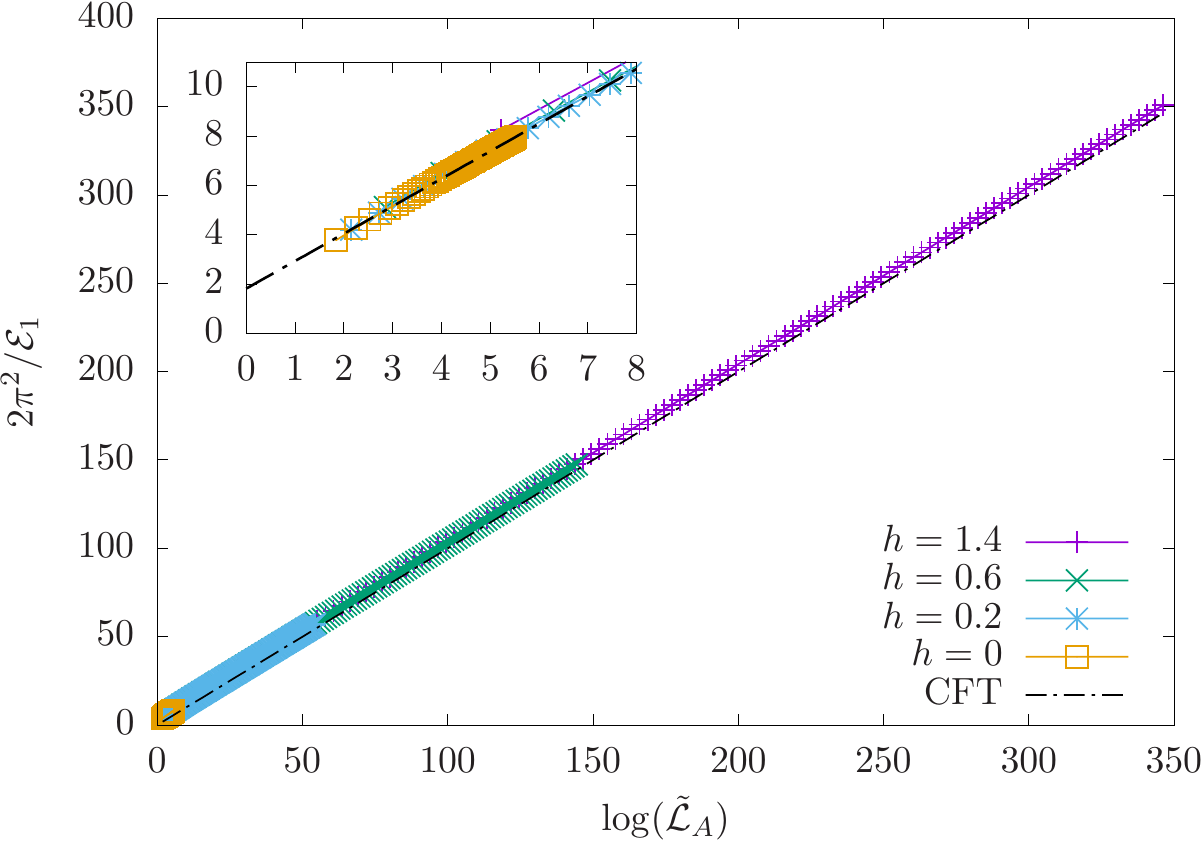}
  \hspace{.1cm}
  \includegraphics[width=.48\textwidth]{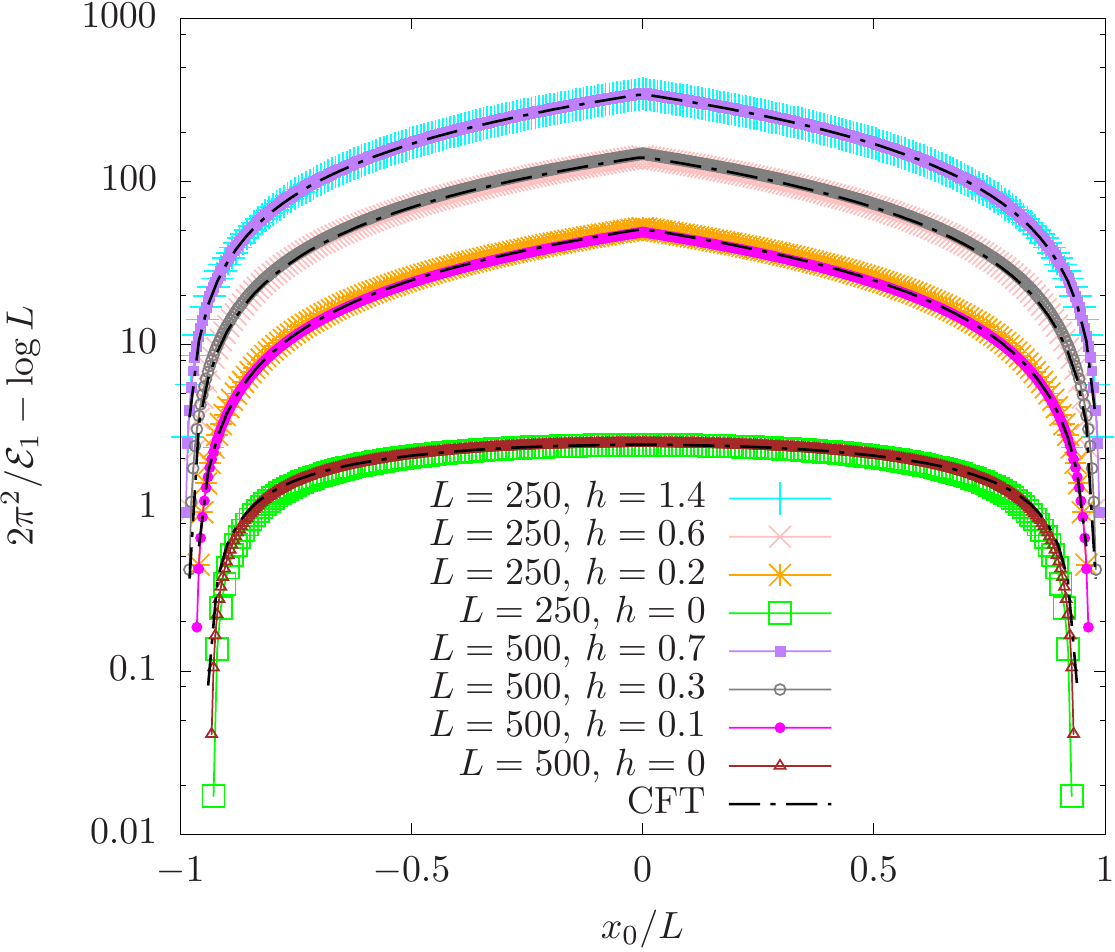}  
\caption{
  Left: The inverse of the first gap in the entanglement spectrum $2\pi^2/{\cal E}_1$ 
  in terms of the logarithm of the effective block size $\log \tilde{\mathcal{L}}_A$, 
  defined in (\ref{tLA}), for $L=250$. 
  The straight dashed-dotted line has slope one, as predicted by the analytic formula (\ref{41a}).
  The inset shows a zoom in the region close to the origin, which highlights the numerical estimation of 
  the intercept, which gives $\log\gamma\approx 1.9$.
  Right: The combination $2\pi^2/{\cal E}_1-\log L$ (see (\ref{41a})) in terms of $x_0/L$ for rainbow chains having either $L=250$ or $L=500$, 
  with values of $h$ properly chosen in order to emphasize the collapse of the numerical data having the same $\lambda=hL$. 
  By employing the same value of $\gamma$ obtained numerically in the left panel,
  an excellent agreement is observed with the theoretical prediction \eqref{eq:tildeLA}.
  }
\label{fig:egap}
\end{figure}

The CFT analysis in \S\ref{sec:EHI} predicts that ${\cal E}_1$ is
given by \eqref{eek tilded} specified to our case, where  $\Delta_1 =1$ and the effective
length (\ref{tLA}) must be employed.
The result reads 
\beq
{\cal E}_1 = \frac{2 \pi^2}{ \log(\gamma \tilde{\mathcal{L}}_A )}\,,
\label{41} 
\eeq
which tells us that ${\cal E}_1$ is inversely proportional to the logarithm of the effective length (\ref{tLA})
and we have introduced a non universal constant $\gamma$ that can be estimated through the numerical analysis. 
We find it more convenient to write \eqref{41} as follows
\beq
  \label{41a} 
\frac{ 2 \pi^2}{{\cal E}_1}  
   \,=\, \log  \tilde{\mathcal{L}}_A  + \log \gamma  
   \,=\, \log L  + \log  \tilde{L}_A + \log \gamma\,.
\eeq

In Fig.\,\ref{fig:egap} we show the numerical data collected for various rainbow chains
in order to check the validity of the analytic prediction (\ref{41a}), where $\tilde{L}_A$
is given by (\ref{eq:tildeLA}).
The agreement with the data point is excellent. 
The constant $\gamma$ in (\ref{41a}) is non universal and it has been estimated through a 
global fit of the data shown in the left panel of Fig.\,\ref{fig:egap}, 
finding $\log \gamma \approx 1.9$ (see the inset).
This numerical value agrees
with the constant $-\,\psi(1/2) \simeq 1.963$, where $\psi(x)$ is the
digamma function, obtained in 
\cite{calabrese-essler}\footnote{We thank Pasquale Calabrese for drawing our attention on \cite{calabrese-essler}.} 
in the analysis of the subleading correction 
to the R\'enyi entropies of an interval in the infinite XX chain 
in the limit $n \to + \infty$ of the R\'enyi index.
It would be interesting to provide a derivation of the constant $\gamma$
through analytical techniques. 

In the right panel of Fig. \ref{fig:egap} 
we have considered the combination $2\pi^2/{\cal E}_1-\log L$  in terms of $x/L$
for rainbow chains of two different lengths, 
showing that the data points collapse on the analytic expression 
given by $ \log  \tilde{L}_A + \log \gamma$,
which corresponds to a family of curves parameterised by $\lambda=hL$, 
as one can observe from (\ref{eq:tildeLA}).

We find it worth discussing the  expression \eqref{41a} in some interesting special regimes. 

In the homogenous case, i.e. for $h=0$, the following result of \cite{Cardy-Tonni16} is recovered
\beq
\label{gap-hom2}
\frac{2\pi^2}{\mathcal{E}_1} \,=\,
  \log \left[ L \, \sin \bigg(\frac{\pi (L - x_0)}{2L}\bigg) \right] 
 + \log (8 \gamma /\pi )\,,
\eeq
which corresponds to the $j=1$ case of \eqref{12} and tells us that
the gap $\mathcal{E}_1$ is inversely proportional to the logarithm of
the chord length of the block.

When $A$ is half of the entire segment, $x_0 =0$ and the
effective length (\ref{tLA}) simplifies to $\tilde{\mathcal{L}}_A =
8\tilde{L}/\pi$; therefore \eqref{41a} becomes
\beq
\frac{2\pi^2}{\mathcal{E}_1} 
\,=\, \log L + \log \left( \frac{e^{\lambda} - 1}{\lambda} \right) + \log (8 \gamma /\pi )\,,
\eeq
which agrees with the ansatz made in Eq.\,(21) of \cite{Ramirez.14b}\footnote{Comparing the notations, we have that $\Delta_L$, $z$  and $6 \,  \tilde{d}(z)$ in \cite{Ramirez.14b} 
correspond respectively to ${\cal E}_1$, $\lambda$  and $\log [ (e^{\lambda} -1)/\lambda] + \log(8 \gamma/\pi)$ in this manuscript.}.

Finally, in the regime defined by $\lambda \gg 1$ with $|x_0|/L$ fixed, we have that \eqref{41a} gives
\beq
\frac{2\pi^2}{\mathcal{E}_1} \, \simeq \, 
h\big(L - |x_0|\big) - \log h +  \log (8 \gamma / \pi) \,.
\eeq
In particular,  for $x_0=0$  the r.h.s. of this expression further simplifies to ${\cal E}_1 \simeq 2 \pi^2/(h L)$, 
which is the result found in \cite{Ramirez.15} and employed to interpret the ground state of the rainbow chain as a thermofield double.

\subsection{A contour for the entanglement entropies}

Entanglement contours are natural concepts to study in the analysis of the bipartite entanglement.
These quantities attempt to quantify the contribution of a single site (or of a point in the continuum)
in the subsystem to the entanglement of the bipartition. 
The contours for the entanglement entropies in some free and homogeneous models on the lattice
have been studied in \cite{br-04, chen-vidal, frerot-roschilde, cdt-17}.

In the strong coupling limit of the rainbow chain, the question addressed by the contour
for the entanglement entropy has a natural answer.
Indeed, since the ground state in this regime is a valence bond state,
the entanglement entropy can be evaluated by counting the bonds
which are broken by the partition.
In particular, each broken bond provides a contribution of $\log 2$ to the entanglement entropy,
which is obtained by the summing of all these contributions \cite{Vitagliano.10, Ramirez.14b, Ramirez.15}.
Thus, for each site in the block, either it contributes with $\log 2$ to the entanglement  or it does not.

In the following analysis we consider a contour for the entanglement entropies in the rainbow chain 
which can be employed in the entire range of the parameters and captures the expected feature
of the strong coupling regime.

For a free fermion on the lattice, the R\'enyi entropies of a subsystem $A$ are computed from the
the eigenvalues $\{\nu_k\}$ of the correlation matrix
$C_{ij}=\<c^\dagger_i c_j\>$ restricted to the subsystem $A$ (i.e. for $i,j \in A$) as follows
\cite{ent-ham-latt}
\beq
S^{(n)}_A=\,
\frac{1}{1-n} \, 
\sum_k  \log ( \nu_k^n + (1-\nu_k)^n) \,.
\eeq
The entanglement entropy corresponds to the limit $n \to 1$ of this expression.

Denoting by $\{\psi_{k,i}\}$ the components of the normalised eigenvector associated to the eigenvalue $\nu_k$,
it is natural to construct the contour function $s^{(n)}_{A}(i)$ as follows \cite{chen-vidal}
\beq
s^{(n)}_{A}(i)  = {1\over 1-n} 
\sum_k |\psi_{k,i}|^2 \,
\log ( \nu_k^n + (1-\nu_k)^n )\,,
\;\;\qquad\;\;
i\,\in\, A\,.
\label{vidal_contour}
\eeq

Since $\sum_{i \in A} |\psi_{k,i}|^2 = 1$ for every $k$, 
it is straightforward to check that (\ref{vidal_contour})
satisfies  the conditions $\sum_i s^{(n)}_{A}(i) = S^{(n)}_A$.
Moreover, $s^{(n)}_{A}(i) \geqslant 0$ for every $i \in A$. 
These two conditions are minimal requirements for a contour function
for the entanglement entropies.
Other properties have been introduced in \cite{chen-vidal} to reduce the 
large arbitrariness of this construction, but a complete list 
characterising the contour for the entanglement entropies in a unique way is
not known.

In our analysis we have adapted the construction of \cite{chen-vidal} to the rainbow chain.

For homogeneous critical systems in the continuum limit,
the profiles of the contour function for the entanglement entropies have been proposed in \cite{cdt-17} (by employing the CFT analysis of \cite{Cardy-Tonni16})
for a particular class of configurations which includes also the one we are considering,
namely an interval adjacent to the boundary of a segment. 
As for the inhomogeneous critical systems discussed above, 
whose continuum limit is described by a CFT in a curved background, 
a contour function for the entanglement entropies of the configuration of our interest has been proposed \S\ref{sec:contour}:
it is given by (\ref{co3}), where the function $\mathcal{S}_A(x) $ has been defined in (\ref{co3bis}).

By employing the inverse of (\ref{weight function homo}) and (\ref{18}),
we can specify $\mathcal{S}_A(x) $ in (\ref{co3bis}) for the rainbow model, finding 
\beq
\label{contour exp}
 \mathcal{S}_A(x) 
\;=\;\frac{\pi}{2L}\;
\frac{\lambda}{e^\lambda - 1}
\;\,
\frac{
  e^{\lambda |x/L|}\, 
\cos\big(\tfrac{\pi}{2} \,\tfrac{(e^{\lambda |x_0/L|} - 1) \,\textrm{sign}(x_0/L) }{e^\lambda -1} \big)
  }{
\sin\big(\tfrac{\pi}{2} \,\tfrac{(e^{\lambda |x/L|} - 1) \,\textrm{sign}(x/L) }{e^\lambda -1} \big)
-
\sin\big(\tfrac{\pi}{2} \,\tfrac{(e^{\lambda |x_0/L|} - 1) \,\textrm{sign}(x_0/L) }{e^\lambda -1} \big)
  } \,,
\eeq
which is the inverse of (\ref{ent temp exp}), as also stated by (\ref{co4}).
From (\ref{contour exp}) it is straightforward to observe that $L\, s_A^{(n)}(x)$ is a function of
$x/L$ parameterised by $\lambda$ and $x_0/L$.

The function (\ref{contour exp}) diverges linearly close to the entangling point, i.e. for $x\to x_0^+$, 
as already remarked through the more general expansion (\ref{contour linear x0}),
and it gets a finite value at the other endpoint, which coincides with the endpoint of the segment.

\begin{figure}[t!]
\vspace{.2cm}
\hspace{-1cm}
  \includegraphics[width=.53\textwidth]{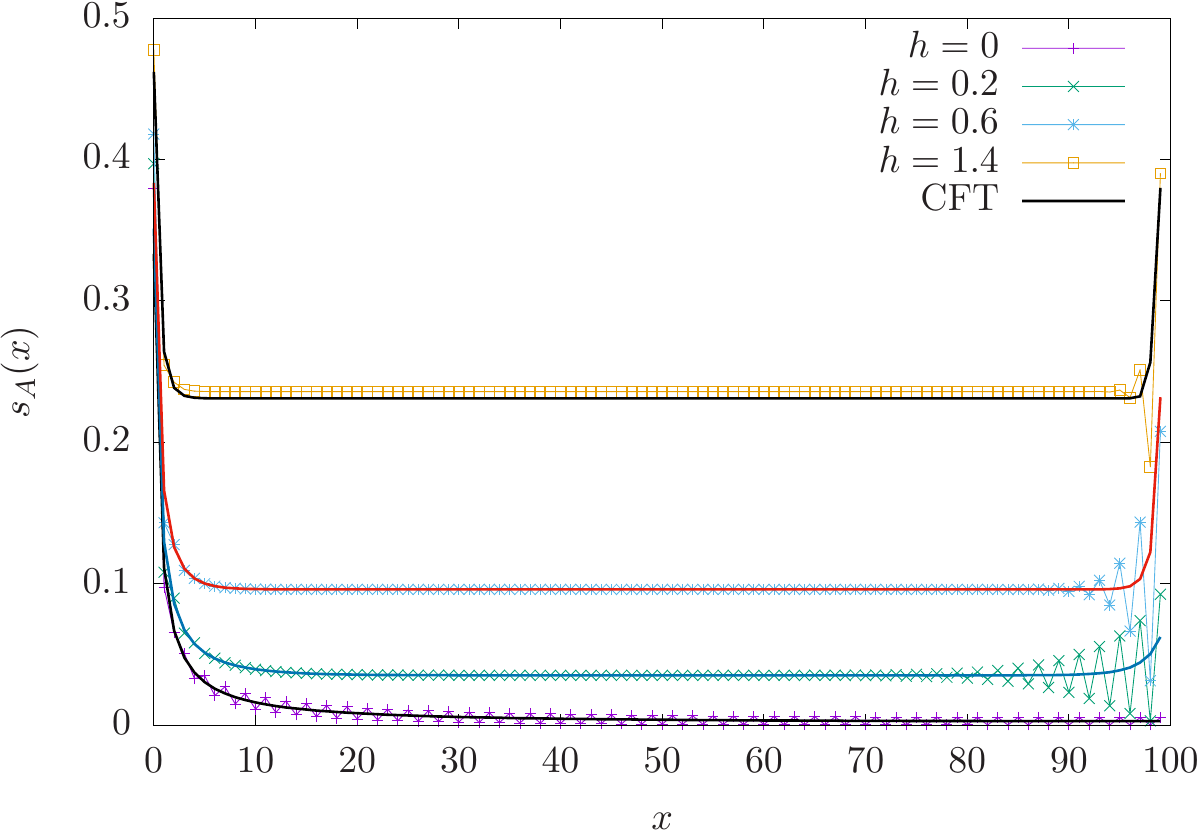}
   \hspace{.2cm}
   \includegraphics[width=.52\textwidth]{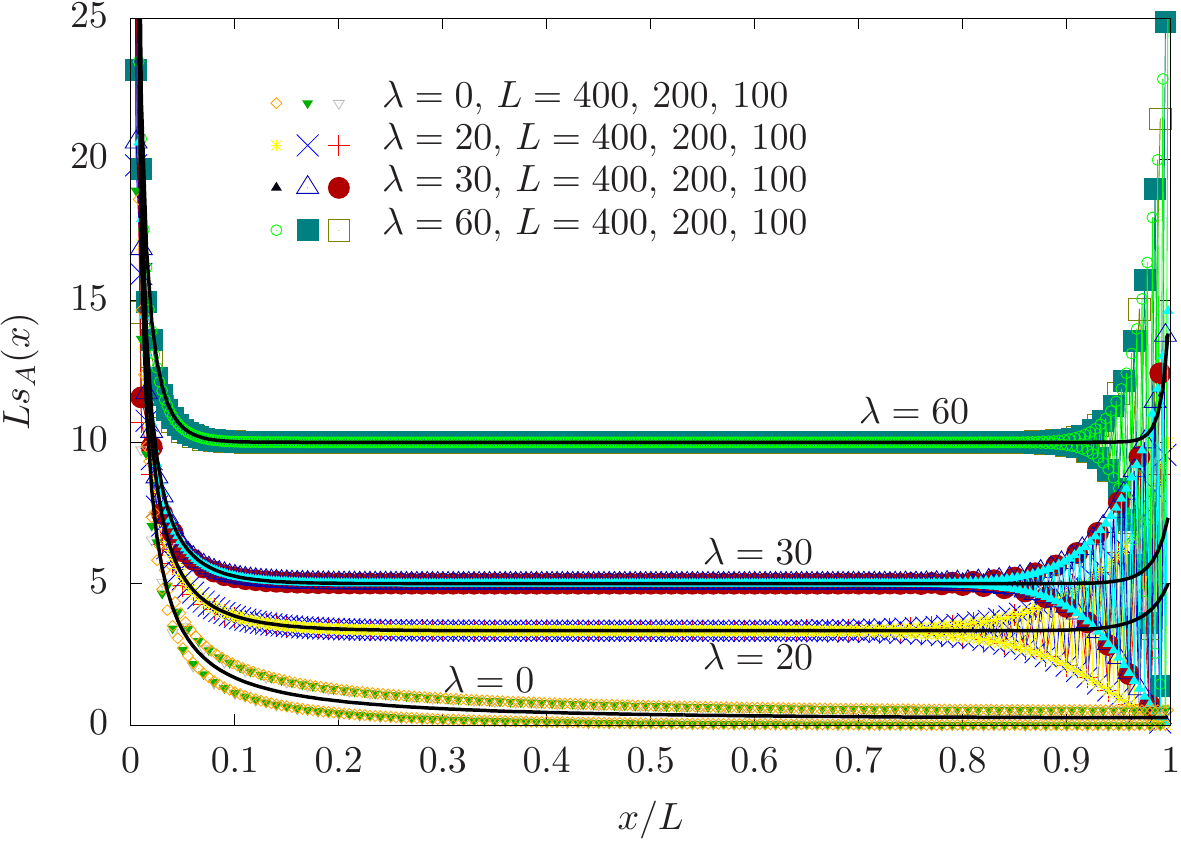}
   \caption{
  Left: Contour function for the entanglement entropy $s_A(x)$ of half rainbow chains with $L=100$ and different values of $h$. 
  The continuous solid lines represent the  CFT prediction given by (\ref{co3}), with $n=1$ and $C_1=0$, and (\ref{81}). 
  A plateau occurs for intermediate values of $x$ and finite $h$, whose height is $h/6$.
  Close to the entangling point the linear divergence (\ref{contour linear x0}) with $x_0=0$ is observed, while near the other endpoint (which is also the endpoint of the segment) non universal parity oscillations become relevant.
  Right:  $L\,s_A(x)$ as a function of $x/L$ for different values of
  $L$ and $h$ suitably chosen in order to deal with the fixed values of $\lambda=hL$ reported near the corresponding curves. 
  }
  \label{fig:cont}
\end{figure}

 We find it worth considering first the contour function for the entanglement entropies given by (\ref{co3}) and (\ref{contour exp})
 in some special regimes. 
 We have set $C_n = 0$ throughout our numerical analysis.

In the homogenous case ($h=0$), we have that $\tilde{x} = x$;
therefore $\tilde{L} = L$ and $\tilde{x}_0 = x_0$.  
In this regime the function (\ref{contour exp}) simplifies to the expression found in \cite{Cardy-Tonni16}, namely
\beq
\label{contour rb func hom}
\mathcal{S}_A(x) 
=
\frac{\pi}{2L} \;
\frac{\cos(\pi x_0 / (2L) ) }{\sin(\pi x/(2L) ) - \sin(\pi x_0 / (2L) )}  \,.
\eeq
The contour function for the entanglement entropies of  this configuration 
has been also studied in \cite{cdt-17} in the homogeneous harmonic chain with Dirichlet boundary conditions imposed at the
endpoints of the segment. 
For this harmonic chain 
the CFT prediction (\ref{contour rb func hom}) 
(which does not depend on the specific boundary conditions imposed at the endpoints of the segment)
does not capture the lattice data for the entire interval.
In particular,
a good agreement with the data points is observed near the entangling point (i.e. for $x\simeq x_0^+$) 
but a large deviation from (\ref{contour rb func hom}) occurs as $x$ approaches the boundary, 
i.e. where the data points are sensible to the boundary condition imposed at the endpoint of the segment.

Another interesting special  case corresponds to a block $A$ given by half of the entire segment. 
In this case $x_0 = 0$ and (\ref{contour exp}) becomes
\beq
\label{81}
\mathcal{S}_A(x) 
=
\frac{\pi h\, e^{h x}}{2(e^{hL}-1)} \;
\bigg[\sin\left( \frac{\pi}{2}   \frac{e^{h x} -1}{e^{hL}-1} \right) \bigg]^{-1}  ,
\eeq
where $x \in (0,L)$.
In the regime of $hL \gg 1$ with $x/L$ kept constant, 
the argument of the sine function becomes $\tfrac{\pi}{2} \, e^{-h(L-x)}$,
therefore the expression in (\ref{81}) simplifies to $\mathcal{S}_A(x) \simeq h$. 
In the homogenous case, i.e. for $h=0$, the function (\ref{81}) simplifies further to
\beq
\mathcal{S}_A(x) 
=
\frac{\pi}{2L \, \sin(\pi x / (2L))} \,,
\eeq
which can be obtained also by setting $x_0$ into (\ref{contour rb func hom}).

The results of our numerical analysis for the contour of the entanglement entropy when $x_0 = 0$
have been reported in Fig.\,\ref{fig:cont}, where the left panel shows the contour function $s_A(x)$ for $L=100$
and different values of $h$.
The solid lines correspond to the CFT prediction given by (\ref{co3}) for $n=1$ and (\ref{81}) with $C_1 = 0$.
The agreement between the numerical data and the analytic expression in the continuum is remarkably good
already at this value of $L$.
In the right panel of Fig.\,\ref{fig:cont} we consider rainbow chains of different lengths and various values of $h$.
As the CFT prediction (\ref{81}) suggests, it is convenient to plot $L\,s_A(x)$ in terms of $x/L$
because the resulting curve is parameterised only by $\lambda$.
This prediction is confirmed by the numerical data shown in the right panel of Fig.\,\ref{fig:cont}.

\begin{figure}[t!]
\vspace{.2cm}
\hspace{-1cm}
\includegraphics[width=.54\textwidth]{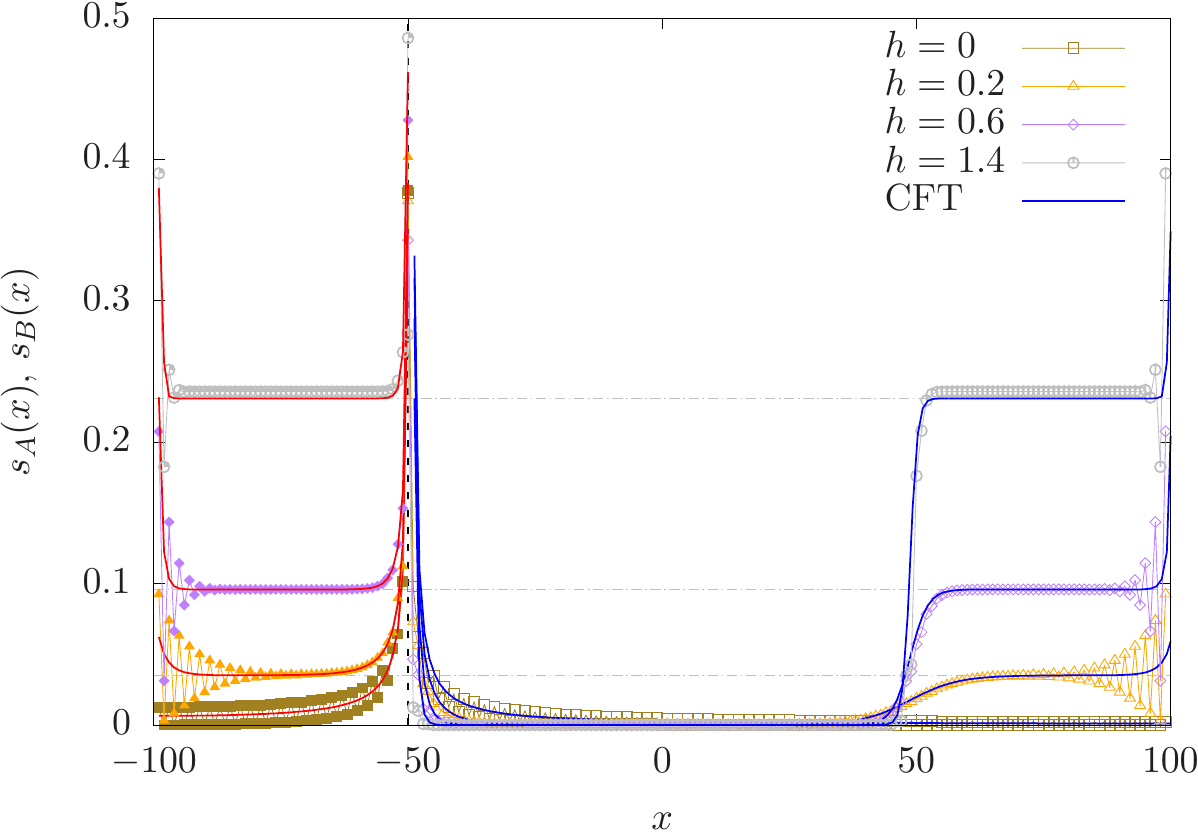}
\hspace{-.0cm}
\includegraphics[width=.53\textwidth]{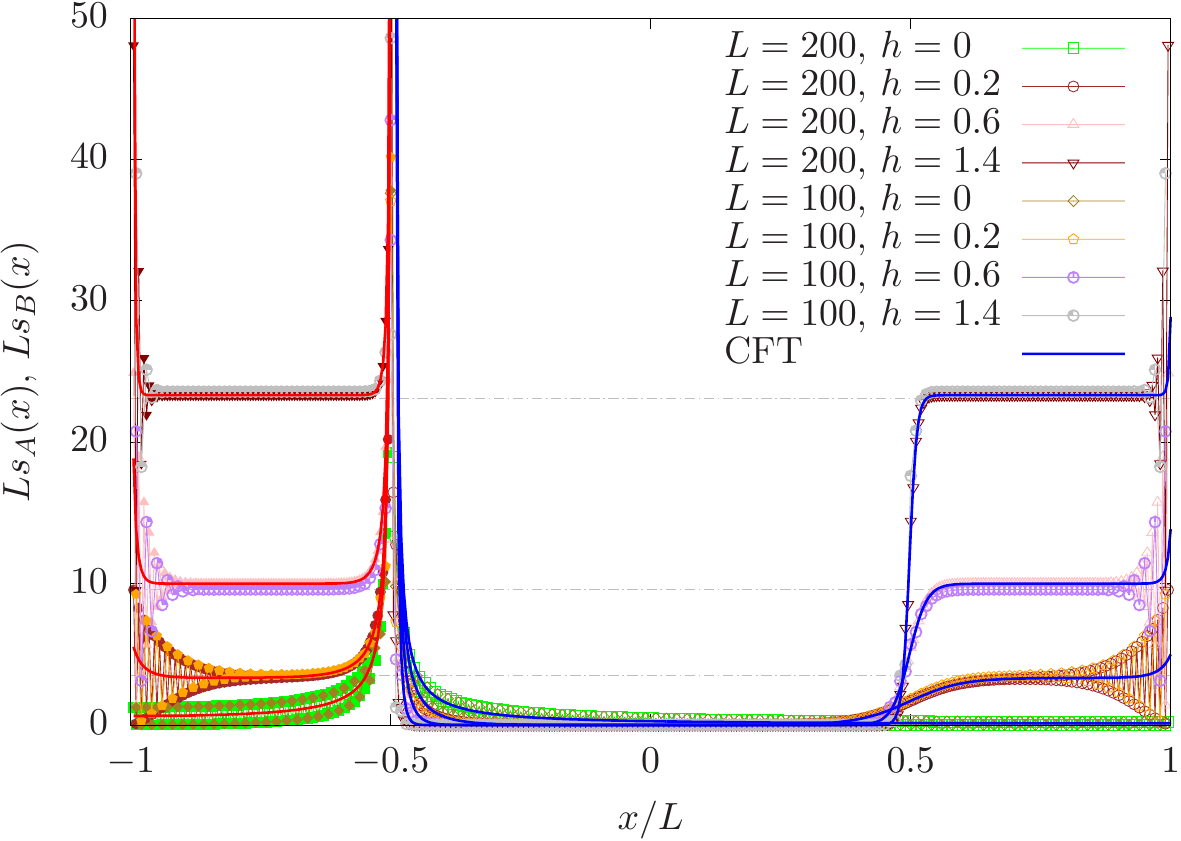}
  \caption{
  Left: Contour functions for the entanglement entropy $s_A(x)$ and $s_B(x)$
  of two complementary blocks $A=(x_0,L)$ and $B=(-L,x_0)$ separated by one entangling point at $x_0 <0$.
  The rainbow chains have $L=100$, $x_0=-50$ and different values of $h$.
  The blue and red solid lines correspond to the CFT prediction for $A$ and $B$ respectively,
  given by (\ref{co3}), with $n=1$ and $C_1 =0$, and (\ref{contour exp}). 
  The linear divergence (\ref{contour linear x0}) is observed in the neighbourhood of the entangling point,
  while finite parity oscillations occur close to the boundaries of the segment. 
  The dashed horizontal lines correspond to $h/6$, which is the height of the plateau predicted by the analytic formula.
  In the regime $\lambda \gg 1$ the contour function vanishes exponentially in the inactive region $(x_0,-x_0)$, 
  as found analytically in (\ref{bond x0neg}),
  confirming the expectation that this region does not contribute to the entanglement between $A$ and $B$ because
  the corresponding bonds do not cross the entangling boundary. 
  Parity oscillations occur also for low non vanishing values of $h$ near the physical boundaries. 
  Right:  $L\,s_A(x)$ and $L\,s_B(x)$ in terms of $x/L$ for two different values of $L$, with $h$ suitably chosen in order to deal with fixed values of $\lambda=hL$.  
}
  \label{fig:cont x0neg}
\end{figure}

The linear divergence (\ref{contour linear x0}) close to the entangling point,
which is the universal prediction of the CFT analysis, is clearly observed from the numerical data.
As for the behaviour of the contour function for the entanglement entropy close to the other endpoint
of the interval $A$, which coincides with the endpoint of the segment, 
the finite value of $s_A(x)$ is captured by the data points within some range established by
non universal parity oscillations. 
From the right panel of Fig.\,\ref{fig:cont}  we observe that these oscillations provide a definite profile under scaling.
It would be interesting to have some analytical comprehension of these oscillations.

\begin{figure}[t!]
\vspace{.2cm}
\hspace{-1.3cm}
\includegraphics[width=.54\textwidth]{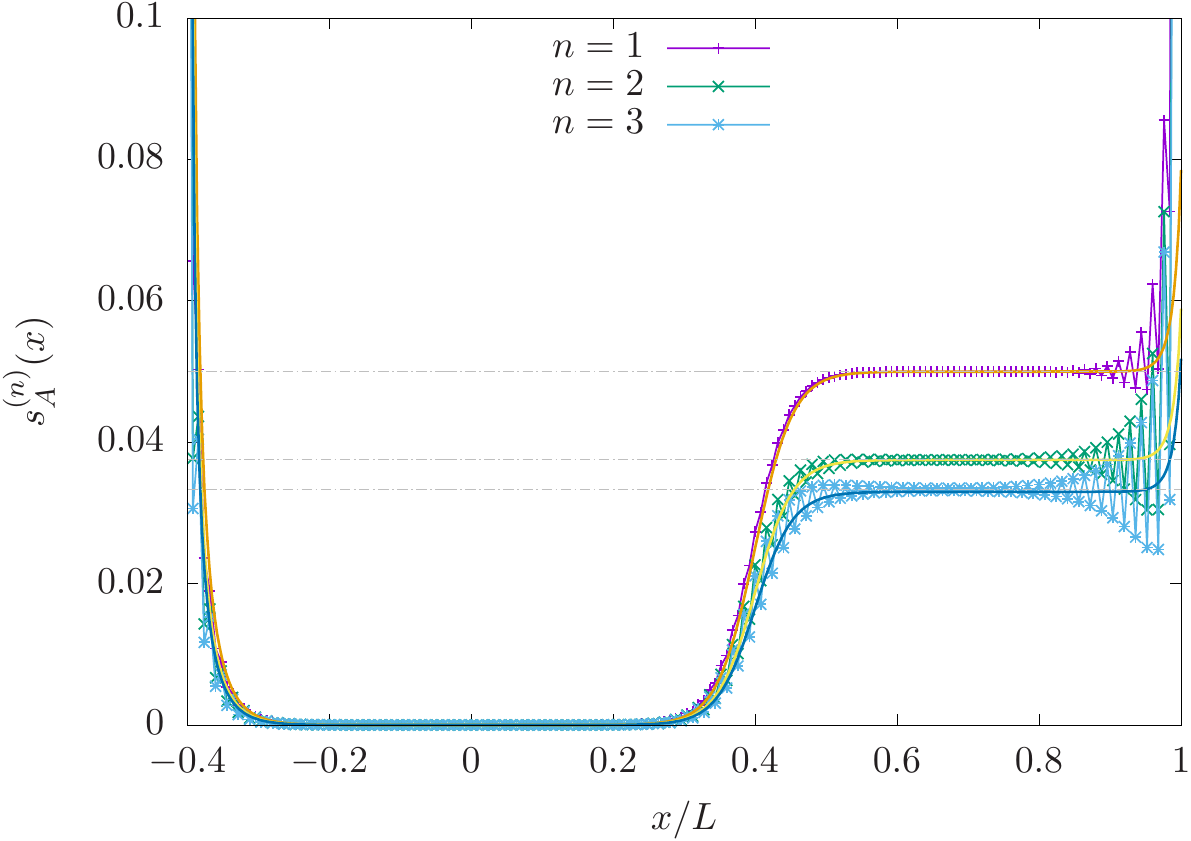}
\hspace{.2cm}
\includegraphics[width=.53\textwidth]{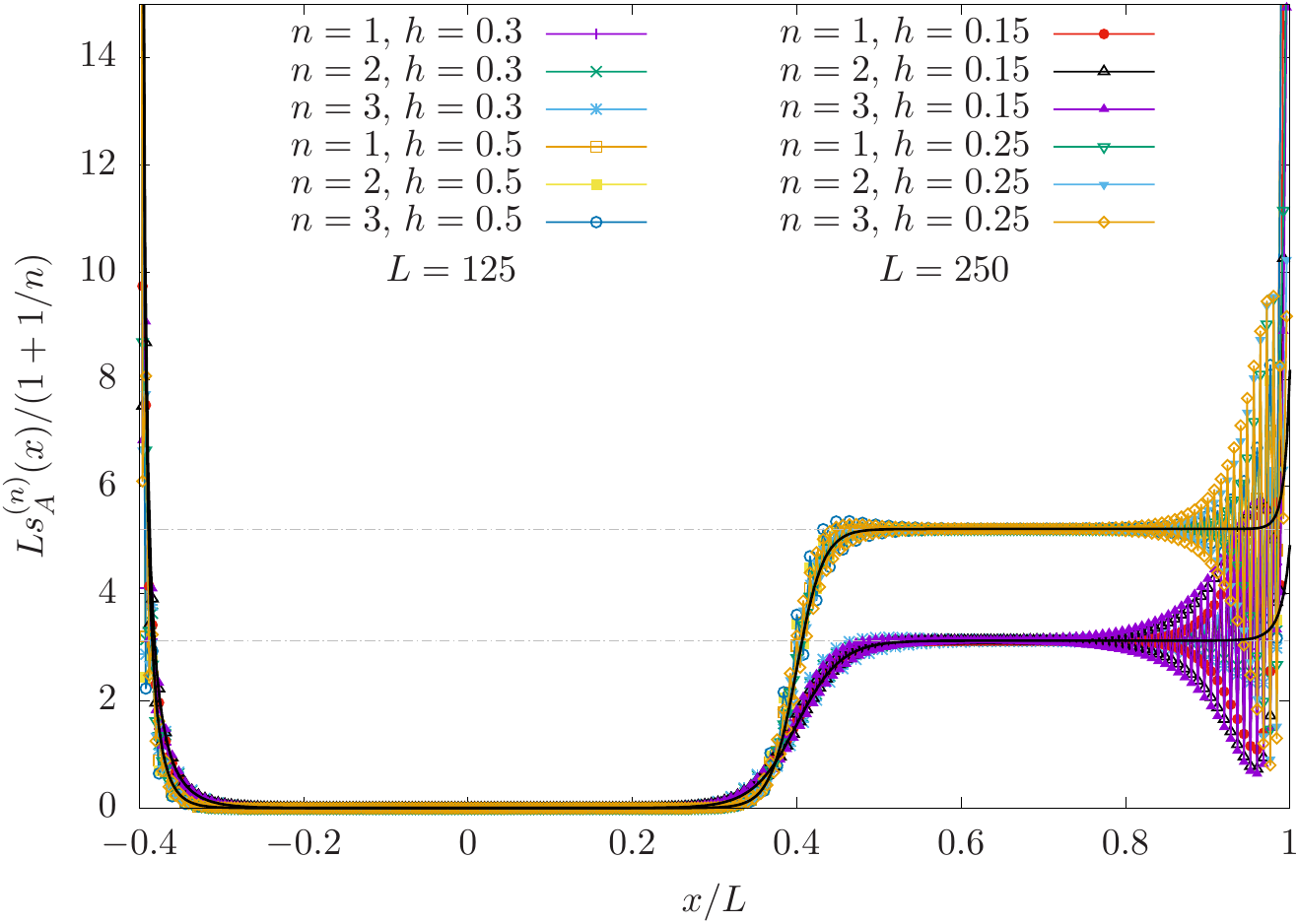}
  \caption{
  Left: 
  Contour function for the entanglement entropies $s_A^{(n)}(x)$
  of an interval $A =(x_0, L)$ adjacent to the boundary, with $x_0 = -50$ and $L=125$. 
  The solid lines represent the CFT prediction  given by (\ref{co3}), with $C_n =0$, and (\ref{contour exp}). 
  The horizontal dashed lines correspond to $(1+\tfrac{1}{n})\,h/12$, which are the heights of the plateaux obtained from the analytic formula.  
  The amplitude of the parity oscillations increases as the R\'enyi index $n$ increases. 
  Right: $L\,s^{(n)}_A(x)/(1+\tfrac{1}{n})$ as a function of $x/L$ for two different system sizes and $x_0 /L = - 0.4$, with $h$ properly chosen in order to emphasize 
  the collapse of the numerical data having the same $\lambda=hL$. 
  }
  \label{fig:cont x0neg renyi}
\end{figure}

The most interesting behaviour for the contour function of the entanglement entropies is observed when $x_0 \neq 0$.
In this case the CFT analytic formula is given by (\ref{co3}) with $C_n = 0$
and (\ref{contour exp}).
The essential feature is captured in the regime of $\lambda \gg 1$ with $x/L$ constant and $x \neq x_0$, 
where (\ref{contour exp}) becomes
\beq
\label{contour rb large}
\mathcal{S}_A(x) 
\, \simeq \,
\frac{h}{\textrm{sign}(x)  - \textrm{sign}(x_0) \,e^{-h(|x|- |x_0|)}  }  \,.
\eeq

When $x_0 > 0$, this expression further simplifies to $\mathcal{S}_A(x) \simeq h$, 
which is the same plateau already found for $x_0 = 0$.

The intriguing behaviour is observed for $x_0 < 0$, where (\ref{contour rb large}) gives
\beq
\label{bond x0neg}
\mathcal{S}_A(x) 
\,\simeq\,
\left\{\begin{array}{ll}
h \,e^{-h(|x_0|- |x|)}  \hspace{1cm}  & -|x_0| < |x| < |x_0|\,,
\\
\rule{0pt}{.6cm}
h&  |x_0| < x < L\,,
\end{array}\right.
\eeq
which is the inverse of (\ref{beta4}), as expected from (\ref{co4}).
We remark that (\ref{bond x0neg}) quantifies the expected feature of the contour function for the entanglement entropies in 
the strong coupling regime mentioned in the beginning of this section (see also the discussion below (\ref{beta4})).
Indeed, since the inactive zone $(-|x_0|,+|x_0|)$ contains bonds which stay inside the block $A$,
it does not contribute to the bipartite entanglement between $A$ and $B$.
Instead, the active zone $(|x_0|,L)$, which contains bonds connecting $A$ to its complement $B$,
is entirely responsible for the entanglement between $A$ and $B$.
Moreover, notice that the plateau profile of the contour function in the active zone $(|x_0|,L)$ of $A$
is the same profile observed for the contour function in $B$ in the intermediate region.

In Fig.\,\ref{fig:cont x0neg} we have collected the data for the contour function of the entanglement entropy 
in various rainbow chains when $x_0 < 0$.
In order to highlight the role of the active zone mentioned above, 
we have shown the contour function for both $A =(x_0, L)$ and its complement $B=(-L, x_0)$.
Notice that, because of the symmetry with respect to the origin, the contour function for $B=(-L, x_0)$ when $x_0 <0$
can be obtained by reflecting the contour function of the block $(-x_0, L)$.
In the left panel of Fig.\,\ref{fig:cont x0neg}, the contour functions $s_A(x)$ and $s_B(x)$ are shown for a fixed
configuration given by  $L=100$ and $x_0 = -50$, 
while in the right panel various configurations have been considered in order to highlight the fact that
$L\,s_A(x)$ and $L\,s_B(x)$ are functions of $x/L$ parameterised by $\lambda$.
In the neighbourhood of the endpoints of the intervals $A$ and $B$, the behaviour is qualitatively like the one discussed
for $x_0$ (see Fig.\,\ref{fig:cont}):
the linear divergence (\ref{contour linear x0}) near the entangling point $x_0$ and finite parity oscillations whose amplitudes increase as $x$
approaches the boundaries of the segment. 
The new feature of Fig.\,\ref{fig:cont x0neg} is the behaviour of the contour function in the neighbourhood of the point
$-x_0$ in $A$, which separates the active zone from the inactive zone.
A remarkable agreement is observed with the CFT analytic formula given by (\ref{co3}) with $C_n = 0$
and (\ref{contour exp}) for both the contour functions $s_A(x)$ and $s_B(x)$.
Notice that, close to the boundaries of the segment, the CFT curve is in the middle of the oscillations. 
We find vey remarkable that the CFT expression for the contour function is able to capture the behaviour of the numerical data for these free fermionic chains 
along all the entire block. 
This does not happen for the harmonic chains \cite{cdt-17}.

\begin{figure}[t!]
\vspace{.2cm}
  \begin{center}
  \includegraphics[width=.65\textwidth]{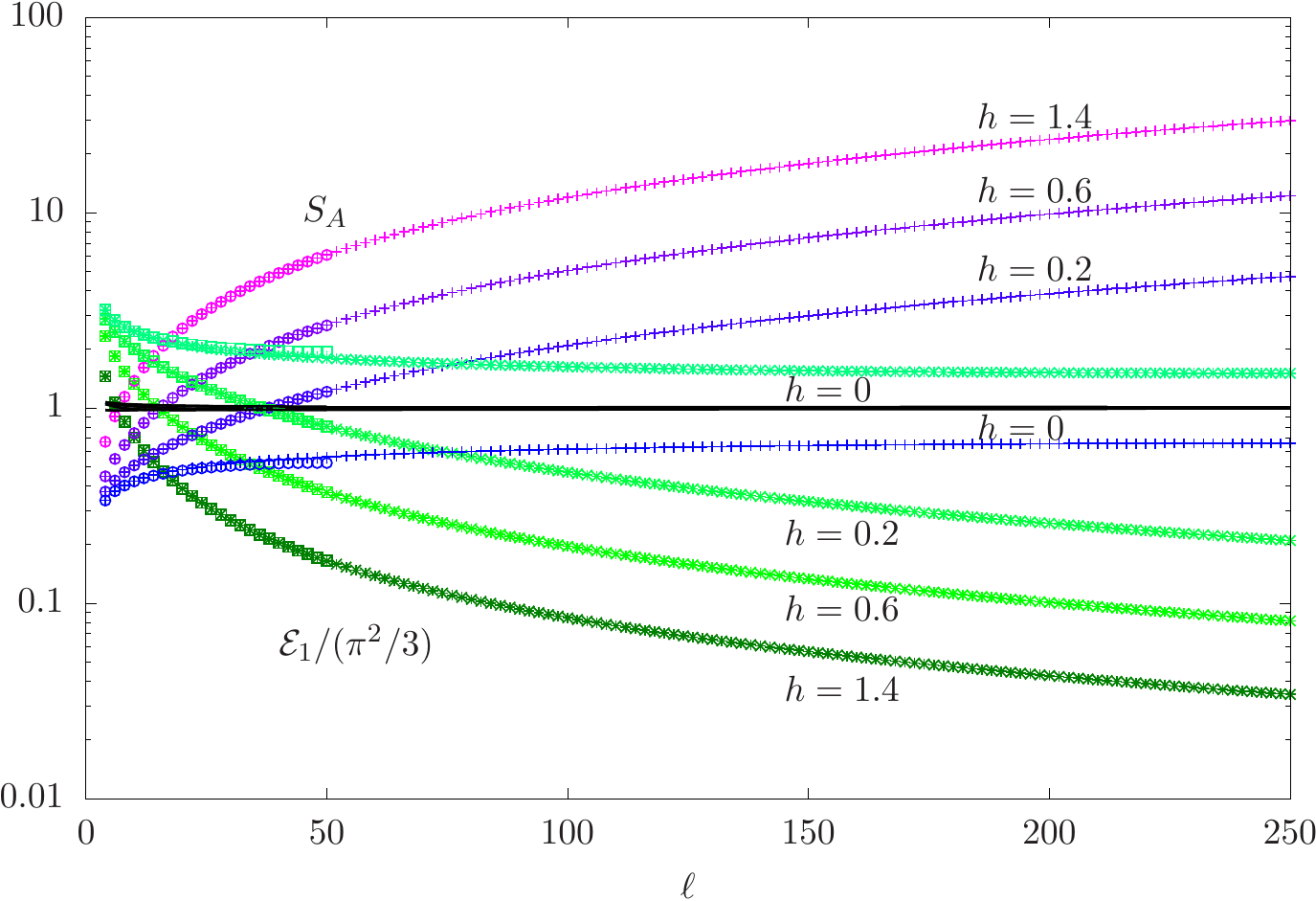}
  \end{center} 
  \vspace{-.3cm}
  \caption{Entanglement entropy $S_A$ (increasing bluish points) and
    first gap in entanglement spectrum ${\cal E}_1/ \tfrac{\pi^2}{3}$ 
    (decreasing greenish points), along with their product (black points).
    The data points confirm the validity of (\ref{consistency}) with $n=1$. 
    Rainbow chains with either $L=50$ or $L=250$ have been employed, being 
    $A=(x_0,L)$ of size $\ell$.
    Notice the symmetry between $S_A$ and ${\cal E}_1/ \tfrac{\pi^2}{3}$ (in logarithmic scale) for the same chain.  
    }
  \label{fig:SG}
\end{figure}

Considering again an interval $A=(x_0, L)$ with $x_0 <0$,
in  Fig.\,\ref{fig:cont x0neg renyi} we have reported the contour function $s_A^{(n)}(x)$ for the R\'enyi entropies $S_A^{(n)}$, 
in order to study the role of the R\'enyi index $n$.
The essential features discussed for the contour function $s_A(x)$ in Fig.\,\ref{fig:cont x0neg} are observed also for 
$s_A^{(n)}(x)$ with $n \geqslant 2$.
We find it worth remarking that, from the data points we can clearly observe that the amplitude of the parity oscillations increases 
as the R\'enyi index $n$ increases. 
Moreover, in the right panel of Fig.\,\ref{fig:cont x0neg renyi} we show that,
by considering $L\, s_A^{(n)}(x) /(1+ \tfrac{1}{n})$, the data points for different rainbow chains collapse on a function of $x/L$ parameterised by $\lambda$ and $x_0/L$.
The CFT prediction for this function can be read from  (\ref{contour exp}) and it nicely agrees with the lattice data.

\begin{figure}[t!]
\vspace{.2cm}
\hspace{-1cm}
  \includegraphics[width=.52\textwidth]{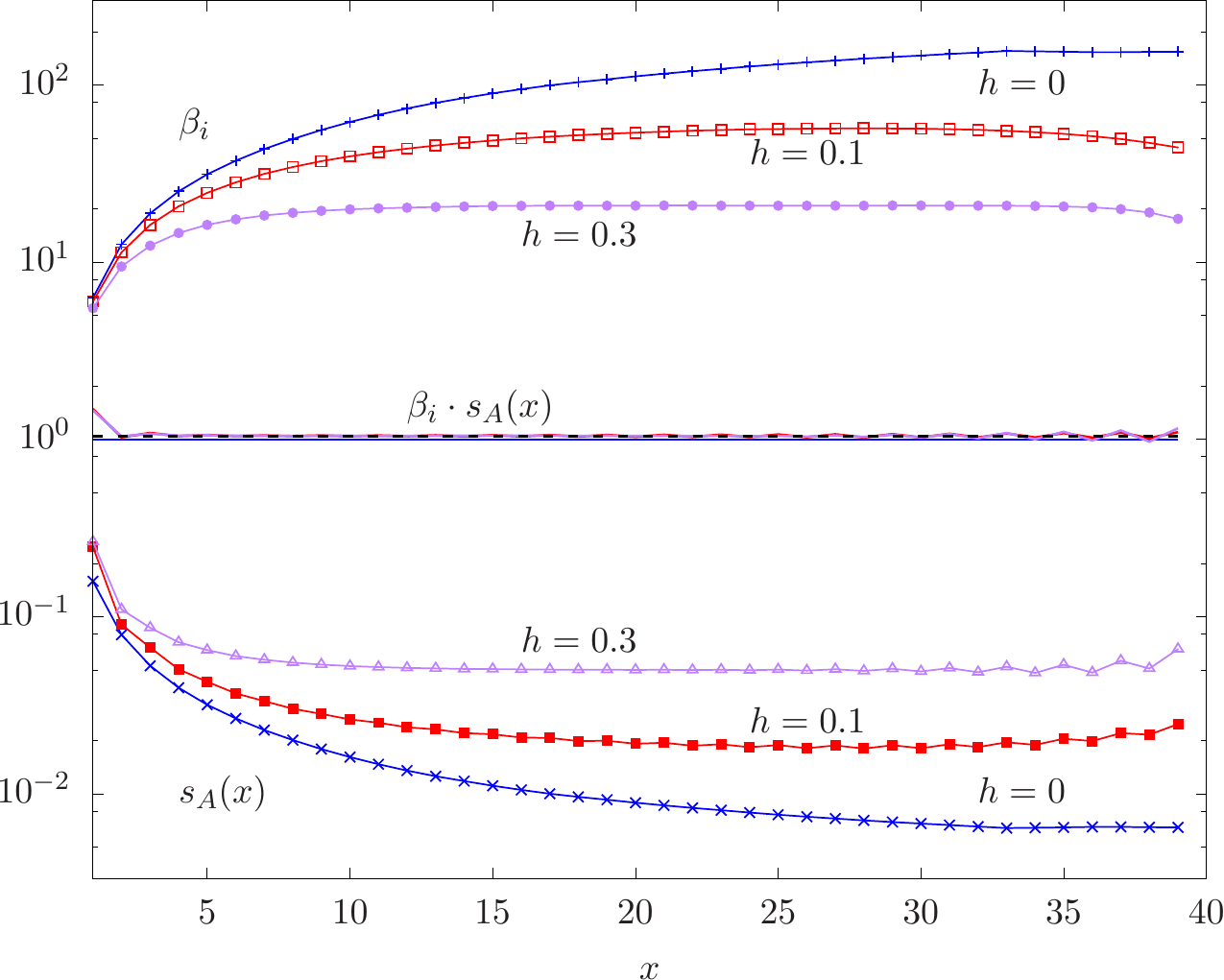}
  \hspace{.3cm}
  \includegraphics[width=.52\textwidth]{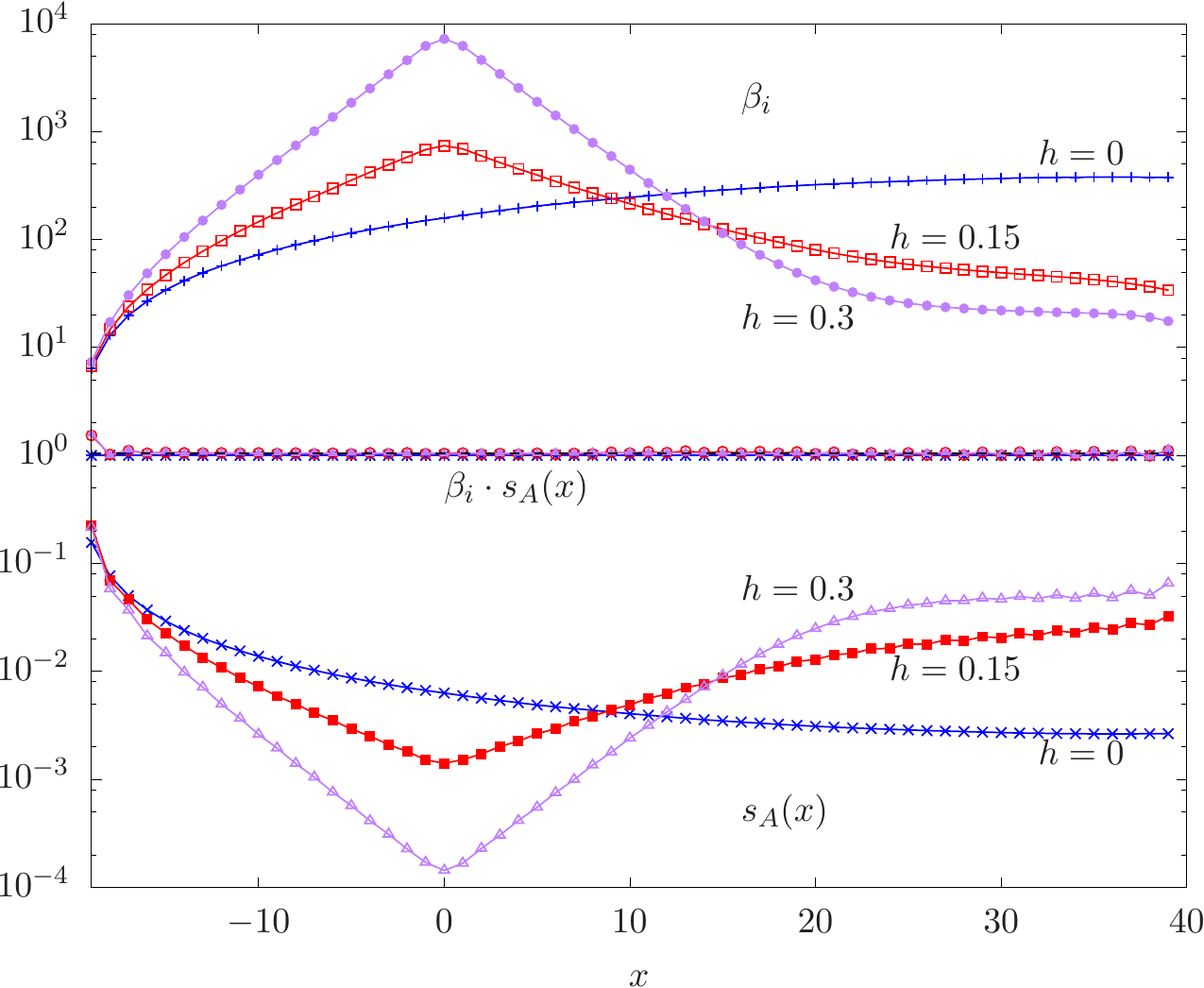}
  \caption{
  Couplings $\beta_i$ of the entanglement hamiltonian (\ref{eq:entham}) and 
  contour function for the entanglement entropy $s_A(x)$ in terms of the position
  inside the subsystem ($i=x$), for blocks $A= (x_0,L)$ with $L=40$, along with their product.
  The logarithmic scale  highlights the validity of the relation $\beta_i\,s_A(x)=\pi/3$ (see (\ref{consistency-local}) with $n=1$).
    The parity oscillations in the contour have been removed by applying a local
    smoothing convolution. 
    Left: Configurations with $x_0=0$.
    Right: Configurations with $x_0/L=-0.5$.}
  \label{fig:sbeta_half}
\end{figure}

In the final part of our numerical analysis of the rainbow chain, 
we find it worth considering the CFT relation (\ref{consist rel1 tilded})
involving the R\'enyi entropies $S_A^{(n)}$ and the gaps $\mathcal{E}_k$ in the entanglement spectrum,
In particular, focussing on the first gap $\mathcal{E}_1$, which has been studied numerically in \S\ref{sec:gap1rb} for the rainbow chain, 
from (\ref{consist rel1 tilded}) we have
\beq
\label{consistency}
\mathcal{E}_1 \, S^{(n)}_A 
\,\simeq\,
\frac{\pi^2}{6} \left( 1 + \frac{1}{n} \right),
\eeq
where the subleading corrections have been neglected. 
In Fig.\,\ref{fig:SG} we show numerical data to check (\ref{consistency}) for $n=1$ in various rainbow chains
(see also Fig.\,8 of \cite{Ramirez.14b}).

Another interesting CFT relation that we find worth checking numerically is (\ref{co5}).
For rainbow chains $c=1$; therefore  (\ref{co5}) becomes
\beq
\label{consistency-local}
\big[ 2\pi \beta_A(x) \big]\, s^{(n)}_A(x) 
\,\simeq\,
\frac{\pi}{6} \left( 1 + \frac{1}{n} \right),
\eeq
up to subleading corrections. 
In Fig.\,\ref{fig:sbeta_half} different rainbow chains have been considered 
to check the relation (\ref{consistency-local}) numerically and a good agreement is observed.


\section{Conclusions}
\label{sec:conclusions}

The analysis of operators and other related quantities underlying the measures of the 
bipartite entanglement is an important task in order to improve our comprehension 
of entanglement in many body quantum systems and quantum field theory.

In this manuscript we have considered some inhomogeneous static critical systems in one spatial dimension
whose continuum limit is described by a CFT in a static curved background characterised by a metric
which is a Weyl rescaling of the flat metric that only depends on the space variable. 
In these models, we have considered a subsystem given by an interval adjacent to the boundary of a segment with finite length.
The same boundary condition is imposed at the endpoints of the segment. 
For this configuration we have studied the entanglement hamiltonian, the entanglement spectrum 
and an entanglement contour for the entanglement entropies. 
Our CFT analysis extends some results obtained in \cite{Cardy-Tonni16, cdt-17} in flat spacetime
and it reproduces the entanglement entropies for this configuration obtained 
through the twist fields method in \cite{Dubail.17,Laguna.17}.

In order to check numerically our CFT predictions through lattice computations, 
we have considered the rainbow chain, which is a free fermion model in a segment 
where the hopping amplitudes decay exponentially going from the center of the chain 
towards the boundaries in a symmetric way \cite{Vitagliano.10,Ramirez.14b,Ramirez.15, Laguna.17}.
Specialising our CFT formulas to the rainbow model, 
we have obtained an excellent agreement with the lattice data
for all the quantities that we have considered:
the  couplings in the ansatz (\ref{eq:entham}) for the entanglement hamiltonian
(Figs.\,\ref{fig:beta1} and \ref{fig:beta3}),
the largest eigenvalue of the reduced density matrix (Fig.\,\ref{fig:largest}),
the first gap in the entanglement spectrum (Fig.\,\ref{fig:egap})
and the contour function for the entanglement entropies 
(Figs.\,\ref{fig:cont}, \ref{fig:cont x0neg} and \ref{fig:cont x0neg renyi}).

Our numerical analysis of the entanglement hamiltonian is based on the ansatz (\ref{eq:entham}),
which includes only the coupling between nearest neighbours sites in the subsystem.
It would be interesting to consider also couplings between sites at generic distance and, 
given the results of \cite{ent-ham-latt, peschel-eisler-17} for homogeneous free fermions,
we expect that their amplitudes are much smaller than the ones between nearest neighbours. 

As for the contour for the entanglement entropies, we find it remarkable that our 
analytic formulas predict the correct behaviour for the entire interval, 
although they do not depend on the boundary condition imposed at the endpoints of the segment. 
This is not case for the harmonic chain \cite{cdt-17}.
It would be interesting to understand better the role of the boundary conditions 
in the contour function for the entanglement entropies.
Another interesting feature observed in our numerical analysis of 
the contour for the entanglement entropies is given by the parity oscillations,
whose amplitude, which depends both on the position and on the inhomogeneity parameter,
increases near the boundary of the segment. 
Providing an analytic treatment of these parity oscillations could be an interesting question
for future studies. 
Besides the issues related to the specific models, let us remind that 
a complete list of properties which defines the contour for the entanglement entropies
in unique way is not available \cite{chen-vidal, cdt-17}.

Our analysis can be extended in various directions.
For instance, more complicated configurations or curved backgrounds corresponding to other  inhomogeneous critical systems can be considered 
\cite{us-future}.
Moreover, since our analytic formulas can be applied also for inhomogeneous models with arbitrary central charge,
it would be interesting to check numerically their validity for some models having $c\neq 1$.

Another important direction for future studies consists in extending the analysis 
described in this manuscript to inhomogeneous systems defined in 
two or three spatial dimensions.

\section*{Acknowledgements}

We would like to thank Pasquale Calabrese,  Ingo
Peschel, Giovanni Ram\'{\i}rez and Stefan Theisen 
and in particular Viktor Eisler and Luca Tagliacozzo
for useful discussions and comments. 
ET is grateful to the Instituto de F\'{\i}sica Te\'orica, Madrid, for the
financial support and the warm hospitality during the initial stage of this work. 
JRL and GS have been supported by the Grant
No. FIS2015-69167-C2-1-P from the Spanish government, QUITEMAD+
S2013/ICE-2801 from the Madrid regional government. 
We also acknowledge the grant SEV-2016-0597 of the 
  ``Centro de Excelencia Severo Ochoa'' Programme.

\newpage


\begin{appendices}

\section*{Appendices}

\section{R\'enyi entropies and Liouville action}
\label{app:liouville}

In this appendix we describe an argument employed in the discussion reported in the beginning of \S\ref{sec:EHI} 
to obtain (\ref{30}).

Given  a two dimensional manifold $\mathcal{M}$ whose boundary $\partial \mathcal{M}$ 
can be   made of  an arbitrary number of disjoint components, let us denote
by $(\mathcal{M}, g)$ the spacetime defined by introducing a two dimensional euclidean metric $g_{\mu\nu}$ on $\mathcal{M}$.
In the following we consider two spacetimes $(\mathcal{M}, g)$ and $(\mathcal{M}, \hat{g})$ 
whose metrics are related through a Weyl factor, namely $g_{\mu\nu} = e^{2\sigma} \hat{g}_{\mu\nu}$.
The partition functions $\mathcal{Z}[\mathcal{M}, g]$ and $\mathcal{Z}[\mathcal{M}, \hat{g}]$ of a 2D CFT with central charge $c$
defined respectively on $(\mathcal{M}, g)$ and on $(\mathcal{M}, \hat{g})$ 
are related as follows \cite{polyakov-81, alvarez-82}
\be
\label{Z weyl}
\mathcal{Z}[\mathcal{M}, g_{\mu\nu}]
\,= \,
e^{ \frac{c}{6} \,S_L [\,\sigma\,; \,\mathcal{M}, \, \hat{g}_{\mu\nu}\,] }\, \mathcal{Z}[\mathcal{M}, \hat{g}_{\mu\nu}]  \,,
\ee
where $S_L [\,\sigma\,; \,\mathcal{M}, \, \hat{g}_{\mu\nu}\,] $ is the Liouville action on $(\mathcal{M}, \hat{g})$,
which is given by
\bea
\label{liouville-action}
S_L [\,\sigma\,; \,\mathcal{M}, \, \hat{g}_{\mu\nu}\,]
& \,\equiv\, &
\frac{1}{4\pi}
\int_{\mathcal{M}}
\Big(\,
\hat{g}^{\mu\nu} \partial_\mu \sigma\, \partial_\nu \sigma
+ \widehat{R}\,\sigma
+ \mu\,e^{2\sigma}
\Big) \sqrt{\hat{g}}\, d^2x
\nonumber
\\
\rule{0pt}{.7cm}
& &
+
\frac{1}{2\pi}
\int_{\partial\mathcal{M}}
\Big(
 \widehat{K}\,\sigma
+ \tilde{\mu}\,e^{\sigma}
\Big) \sqrt{\hat{h}}\, d\lambda \,,
\eea
being $\widehat{R}$ the scalar curvature of the metric $\hat{g}_{\mu\nu}$.
The metric on $\partial \mathcal{M}$  is the metric induced from the embedding.
The element of arc length on $\partial \mathcal{M}$ is $\sqrt{\hat{h}}\, d\lambda$
and $\widehat{K}$ is the extrinsic curvature of $\partial \mathcal{M}$.
The constants $\mu$ and $\tilde{\mu}$ are respectively 
the bulk cosmological constant and boundary cosmological constant.

The Liouville theory is a paradigmatic model of irrational CFT whose analysis
led to important advances in the comprehension of two dimensional CFTs \cite{ZZ, FZZ, ZZpseudo}.
It is  worth mentioning that the regularization procedure discussed in \S\ref{sec:EHhom} and \S\ref{sec:EHI} following
\cite{H94, Cardy-Tonni16}, which is  based on the removal of infinitesimal disks centered in the positions of the local operators, 
has  been also employed in the path integral approach to the correlation functions in Liouville theory
\cite{ZZ, liouville-geom}.

We are interested in the case of  $\mathcal{M} = \INT \setminus \cup_j \D_{\epsilon}(x_j) $ is a vertical strip $\INT$ 
from which small disks $\D_{\epsilon}(x_j) $ of radius $\epsilon$ centered at the entangling points 
(which can be made by several  disjoint intervals in the segment $(-L,L)$) have been removed. 
We remark that, since  $x_j $ correspond to the entangling points, we have that $x_j \neq -L$ and $x_j \neq L$.
Thus, the boundary $\partial \mathcal{M}$ is the union of the vertical lines given by $x=-L$ and $x=L$, 
and of all the circumferences $\partial \D_{\epsilon}(x_j) $ around  the entangling points $x_j$.

When $\hat{g}_{\mu\nu} = \delta_{\mu\nu} $ is the flat metric, we have $\widehat{R} = 0$ in (\ref{liouville-action}).
As for the extrinsic curvature of $\partial \mathcal{M}$ in the flat background, $\widehat{K} = 0$ along the vertical straight lines
at $x=-L$ and $x=L$, while $\widehat{K} = - 1/\epsilon $ along $\partial \D_{\epsilon}(x_j)$.
Thus, the boundary term containing the extrinsic curvature in (\ref{liouville-action}) provides non vanishing
contributions only along the infinitesimal circumferences $\partial \D_{\epsilon}(x_j)$. For each of them we have
\be
\label{sing_integral K}
\lim_{\epsilon\, \to \,0}\,
\frac{1}{2\pi} \oint_{\partial \D_{\epsilon}(x_j) }  \widehat{K}\,\sigma \, d\lambda
\,=\,
-  \, \sigma(x_j)\,,
\ee
which can be found by assuming that $\sigma$ is smooth in a neighbourhood of the entangling point
and using that $d\lambda = \epsilon \, d\theta$, being $\theta \in [0,2\pi)$ the angular coordinate 
along $\partial \D_{\epsilon}(x_j) $.

In this manuscript we focus on the simplest configuration where $A =(x_0, L)$ is a single interval 
adjacent to the boundary of the strip and 
$\hat{g}_{\mu\nu} = \delta_{\mu\nu} $ is the flat metric.
In this case $\mathcal{M} =  \INT \setminus  \D_{\epsilon}(x_0) $, as discussed in \S\ref{sec:EHhom}.
The spacetime $(\mathcal{M},\delta)$, which corresponds to both $A =(x_0, L)$ and 
its complement $B = (-L, x_0)$, is shown in the left panel of Fig.\,\ref{fig:map}.

In order to find the R\'enyi entropies, we need to consider the following ratio
\be
\label{renyiM}
\Tr\rho_A^n 
\,=\, \frac{\mathcal{Z}[\mathcal{M}_n, g_{\mu\nu}]}{\big(\mathcal{Z}[\mathcal{M}, g_{\mu\nu}]\big)^n}
\,=\,
e^{ \frac{c}{6} ( S_L [\,\sigma\,; \,\mathcal{M}_n, \, \delta_{\mu\nu}\,] -\,n\, S_L [\,\sigma\,; \,\mathcal{M}, \, \delta_{\mu\nu}\,] ) }\
\frac{\mathcal{Z}[\mathcal{M}_n, \delta_{\mu\nu}]}{\big(\mathcal{Z}[\mathcal{M}, \delta_{\mu\nu}]\big)^n}\,,
\ee
where $\mathcal{Z}[\mathcal{M}_n, g_{\mu\nu}]$ is the partition function of the 2D CFT on the $n$-sheeted
Riemann surface obtained by gluing $n$ copies of $\mathcal{M}$ cyclically along the upper and the lower edges of $A$ \cite{H94, Cardy-Tonni16}.
The final expression in (\ref{renyiM}) has been obtained by using (\ref{Z weyl}).
In the case we are considering, $\partial \mathcal{M}_n$ includes $n$ vertical straight lines corresponding to $x=-L$
and $n$ vertical straight lines corresponding to $x=L$ coming from the different copies of $\mathcal{M}$.
Because of the gluing procedure, $\partial \mathcal{M}_n$ also includes a closed
$n$-covering $\partial \D^{n}_{\epsilon}(x_0)$ of the circumference $\partial \D_{\epsilon}(x_0)$.
In particular, $\partial \D^{n}_{\epsilon}(x_0)$ has radius $\epsilon$ and its length is $2\pi n  \epsilon$.
Thus, since $\widehat{K} = -  1/\epsilon$ along $\partial \D^{n}_{\epsilon}(x_0) $, we have
\be
\label{n_integral K}
\lim_{\epsilon\, \to \,0}\,
\frac{1}{2\pi} \oint_{\partial \D^{n}_{\epsilon}(x_0) }  \widehat{K}\,\sigma \, d\lambda
\,=\,
-  \, n\,\sigma(x_0)\,.
\ee

Since in the manuscript we have considered $\sigma = \sigma(x)$ (see (\ref{14})), in the Liouville action (\ref{liouville-action})
the kinetic term, the bulk cosmological term and the boundary cosmological term 
along the straight lines at $x=-L$ and $x=+L$
provide diverging contributions because of the integration in the $y$ direction.
Nonetheless, in (\ref{renyiM}) these divergencies in $S_L [\,\sigma\,; \,\mathcal{M}_n, \, \delta_{\mu\nu}\,] $ cancel with the 
corresponding divergencies coming from the same kind of terms in $S_L [\,\sigma\,; \,\mathcal{M}, \, \delta_{\mu\nu}\,] $. 
Instead, the boundary term in the Liouville action (\ref{liouville-action}) containing 
the extrinsic curvature gives a finite contribution both in $S_L [\,\sigma\,; \,\mathcal{M}_n, \, \delta_{\mu\nu}\,] $
and $S_L [\,\sigma\,; \,\mathcal{M}, \, \delta_{\mu\nu}\,] $.
By employing (\ref{sing_integral K}) and (\ref{n_integral K}) notice that they also simplify in  (\ref{renyiM}). 

The above observations lead to the argument employed in the text below (\ref{29}).

\end{appendices}

\newpage

\end{document}